\documentclass[iop,revtex4]{emulateapj}
\pdfoutput=1  
\usepackage{lscape}
\usepackage{graphicx,natbib,url,twoopt}
\usepackage[varg]{txfonts}
\usepackage{hyperref}               

\bibpunct{(}{)}{;}{a}{}{,}    

\received{receipt date}
\revised{revision date}
\accepted{acceptance date}
\begin{document}  

\title{Red noise versus planetary interpretations in the microlensing event OGLE-2013-BLG-446}
\author{E. Bachelet$^{R0}$,
D. M. Bramich$^{R0}$,
C.~Han$^{K0}$,
J.~Greenhill$^{P0}$,
R. A. Street$^{R1}$,
A.~Gould$^{U0}$,\\
and,\\
G. D'Ago$^{R2,R3}$,
K. AlSubai$^{R0}$,
M. Dominik$^{R4,\footnote{Royal Society University Research Fellow}}$,
R. Figuera Jaimes$^{R4,R5}$,
K. Horne$^{R4}$,
M. Hundertmark$^{R4}$,
N. Kains$^{R5}$,
C. Snodgrass$^{R6,R7}$,
I. A. Steele$^{R8}$,
Y. Tsapras$^{R1,R9}$,
\\
(The RoboNet collaboration)\\
M. D. Albrow$^{P1}$,
V. Batista$^{P2,U0}$,
J.-P. Beaulieu$^{P2}$,
D.P. Bennett$^{P3}$,
S. Brillant$^{P4}$,
J. A. R. Caldwell$^{P5}$, 
A. Cassan$^{P2}$,
A. Cole$^{P0}$,
C. Coutures$^{P2}$,
S. Dieters$^{P0}$,
D. Dominis Prester$^{P6}$, 
J. Donatowicz$^{P7}$, 
P. Fouqu\'e$^{P8,P9}$,
K. Hill$^{P0}$,
J.-B. Marquette$^{P2}$, 
J. Menzies$^{P10}$,
C. Pere$^{P2}$,
C. Ranc$^{P2}$,
J. Wambsganss$^{P11}$,
D. Warren$^{P0}$,\\
(The PLANET collaboration)\\
L. Andrade de Almeida$^{U1}$,
J.-Y. Choi$^{K0}$, 
D. L. DePoy$^{U2}$,
S. Dong$^{U0,U3}$,
L.-W. Hung$^{U0}$,
K.-H. Hwang$^{K0}$,
F. Jablonski$^{U1}$,
Y, K, Jung$^{K0}$, 
S. Kaspi$^{U4}$,
N. Klein$^{U4}$,
C.-U. Lee$^{U5}$,
D. Maoz$^{U4}$,
J.A. Mu\~noz$^{U1}$, 
D. Nataf$^{U0}$,
H. Park$^{K0}$,
R. W. Pogge$^{U0}$, 
D. Polishook$^{U4}$,
I.-G. Shin$^{K0}$,
A. Shporer$^{U6,U7}$,
J.C.Yee$^{U0}$\\
(The $\mu$FUN collaboration)\\
F.~Abe$^{M4}$,
A.~Bhattacharya$^{M1}$,
I.A.~Bond$^{M3}$,
C.S.~Botzler$^{M5}$,
M.~Freeman$^{M5}$,
A.~Fukui$^{M6}$,
Y.~Itow$^{M4}$,
N.~Koshimoto$^{M1}$,
C.H.~Ling$^{M3}$,
K.~Masuda$^{M4}$,
Y.~Matsubara$^{M4}$,
Y.~Muraki$^{M4}$,
K.~Ohnishi$^{M7}$,
L.C.~Philpott$^{M8}$,
N.~Rattenbury$^{M8}$,
To.~Saito$^{M9}$,
D.J.~Sullivan$^{M10}$,
T.~Sumi$^{M0}$,
D.~Suzuki$^{M2}$,
P.,J.~Tristram$^{M9}$,
A. Yonehara$^{M10}$\\
(The MOA collaboration)\\
V.\ Bozza\,$^{m0,m1}$,               
S.\ Calchi Novati\,$^{m2,m0,m3}$,         
S.\ Ciceri\,$^{m4}$,                              
P.\ Galianni\,$^{R4}$,
S.-H.\ Gu\,$^{m5,m6}$, 
K.\ Harps{\o}e\,$^{m7,m8}$,                    
T.\ C.\ Hinse\,$^{m9}$,                            
U.\ G.\ J{\o}rgensen\,$^{m7,m8}$,            
D.\ Juncher\,$^{m7,m8}$, 
H.\ Korhonen\,$^{m10,m7,m8}$, 
L.\ Mancini\,$^{m4}$,
C. Melchiorre,$^{m0,m1}$, 
A.\ Popovas\,$^{m7,m8}$,
A. Postiglione,$^{m15,m16}$,                   
M.\ Rabus\,$^{m11,m4}$,                            
S.\ Rahvar\,$^{m12}$,                              
R.\ W.\ Schmidt\,$^{P11}$, 
G. Scarpetta ,$^{m0,m1,m3}$,
J.\ Skottfelt\,$^{m7,m8}$,                     
John Southworth\,$^{m13}$,
An. Stabile,$^{m0,m1}$,                          
J.\ Surdej\,$^{m14}$, 
X.-B.\ Wang\,$^{m5,m6}$,                    
O.\ Wertz\,$^{m14}$\\
(The MiNDSTEp collaboration)}
\affil{$^{R0}$Qatar Environment and Energy Research Institute, Qatar Foundation, P.O. Box 5825, Doha, Qatar}
\affil{$^{R1}$Las Cumbres Observatory Global Telescope Network, 6740 Cortona Drive, Suite 102, Goleta, CA 93117 USA}
\affil{$^{R2}$Dipartimento di Fisica "E.R. Caianiello", Universit{\`a} di Salerno, Via Ponte Don Melillo, 84084-Fisciano (SA), Italy}
\affil{$^{R3}$Istituto Nazionale di Fisica Nucleare, Sezione di Napoli, Napoli, Italy}
\affil{$^{R4}$SUPA, School of Physics \& Astronomy, University of St Andrews, North Haugh, St Andrews KY16 9SS, UK}
\affil{$^{R5}$European Southern Observatory, Karl-Schwarzschild-Str. 2, 85748 Garching bei M\"unchen, Germany}
\affil{$^{R6}$Max Planck Institute for Solar System Research, Justus-von-Liebig-Weg 3, 37077 Göttingen, Germany}
\affil{$^{R7}$Planetary and Space Sciences, Department of Physical Sciences, The Open University, Milton Keynes, MK7 6AA, UK}
\affil{$^{R8}$Astrophysics Research Institute, Liverpool John Moores University, Liverpool CH41 1LD, UK}
\affil{$^{R9}$School of Physics and Astronomy, Queen Mary University of London, Mile End Road, London E1 4NS, UK}

\affil{$^{P0}$School of Mathematics and Physics, University of Tasmania, Private Bag 37, Hobart, TAS 7001, Australia}
\affil{$^{P1}$Department of Physics and Astronomy, University of Canterbury, Private Bag 4800, Christchurch 8020, New Zealand}
\affil{$^{P2}$UPMC-CNRS, UMR 7095, Institut d’Astrophysique de Paris, 98bis boulevard Arago, F-75014 Paris, France}
\affil{$^{P3}$Department of Physics, 225 Nieuwland Science Hall, University of Notre Dame, Notre Dame, IN 46556, USA}
\affil{$^{P4}$European Southern Observatory, Casilla 19001, Vitacura 19, Santiago, Chile}
\affil{$^{P5}$McDonald Observatory, 16120 St. Hwy Spur 78 \#2, Fort Davis, TX 79734, USA}
\affil{$^{P6}$Department of Physics, University of Rijeka, Omladinska 14, 51000 Rijeka, Croatia}
\affil{$^{P7}$Technische Universitat Wien, Wieder Hauptst. 8-10, A-1040 Vienna, Austria}
\affil{$^{P8}$IRAP, CNRS - Universit\'e de Toulouse, 14 av. E. Belin, F-31400 Toulouse, France}
\affil{$^{P9}$CFHT Corporation 65-1238 Mamalahoa Hwy Kamuela, Hawaii 96743, USA}
\affil{$^{P10}$South African Astronomical Observatory, PO Box 9, Observatory 7935, South Africa}
\affil{$^{P11}$	Astronomisches Rechen-Institut (ARI), Zentrum f\"ur Astronomie der Universit\"at Heidelberg (ZAH), M\"onchhofstr. 12-14, 69120 Heidelberg, Germany}

\affil{$^{U0}$Department of Astronomy, Ohio State University, 140 W. 18th Ave., Columbus, OH 43210,USA}
\affil{$^{U1}$Divisao de Astrofisica, Instituto Nacional de Pesquisas Espaciais, Avenida dos Astronautas, 1758 Sao Jos\'e dos Campos, 12227-010 SP, Brazil}
\affil{$^{U2}$Department of Physics and Astronomy, Texas A\&M University, College Station, TX 77843-4242, USA}
\affil{$^{U3}$Institute for Advanced Study, Einstein Drive, Princeton, NJ 08540, USA}
\affil{$^{U4}$School of Physics and Astronomy and Wise Observatory, Tel-Aviv University, Tel-Aviv 69978, Israel}
\affil{$^{U5}$Korea Astronomy and Space Science Institute, 776 Daedukdae-ro, Yuseong-gu, Daejeon 305-348, Republic of Korea}
\affil{$^{U6}$Las Cumbres Observatory Global Telescope Network, 6740B Cortona Dr, Goleta, CA 93117, USA}
\affil{$^{U7}$Department of Physics, University of California, Santa Barbara, CA 93106, USA}

\affil{$^{K0}$Department of Physics, Chungbuk National University, Cheongju 361-763, Republic of Korea}

\affil{$^{M0}$Department of Earth and Space Science, Graduate School of Science, Osaka University, Toyonaka, Osaka 560-0043, Japan,\\
e-mail: {\tt sumi@ess.sci.osaka-u.ac.jp}}
\affil{$^{M1}$Department of Physics, University of Notre Dame, Notre Dame, IN 46556, USA; bennett@nd.edu}
\affil{$^{M2}$Institute of Information and Mathematical Sciences, Massey University,
Private Bag 102-904, North Shore Mail Centre, Auckland, New Zealand;
i.a.bond,c.h.ling,w.sweatman@massey.ac.nz}
\affil{$^{M3}$Solar-Terrestrial Environment Laboratory, Nagoya University, Nagoya 464-8601, Japan; abe,furusawa,itow,kmasuda,ymatsu@stelab.nagoya-u.ac.jp}
\affil{$^{M4}$Department of Physics, University of Auckland, Private Bag 92019, Auckland, New Zealand; c.botzler,p.yock@auckland.ac.nz}
\affil{$^{M5}$Okayama Astrophysical Observatory, National Astronomical Observatory of Japan, 3037-5 Honjo, Kamogata, Asakuchi, Okayama 719-0232, Japan}
\affil{$^{M6}$Nagano National College of Technology, Nagano 381-8550, Japan}
\affil{$^{M11}$Department of Earth, Ocean and Atmospheric Sciences, University of British Columbia, Vancouver, British Columbia, V6T 1Z4, Canada}
\affil{$^{M7}$Tokyo Metropolitan College of Aeronautics, Tokyo 116-8523, Japan}
\affil{$^{M8}$School of Chemical and Physical Sciences, Victoria University, Wellington, New Zealand}
\affil{$^{M9}$Mt. John University Observatory, P.O. Box 56, Lake Tekapo 8770, New Zealand}
\affil{$^{M10}$Department of Physics, Faculty of Science, Kyoto Sangyo University, 603-8555 Kyoto, Japan}

\affil{$^{m0}$\,Dipartimento di Fisica ``E.R. Caianiello'', Universit\`a di Salerno, Via Giovanni Paolo II 132, 84084, Fisciano (SA), Italy}
\affil{$^{m1}$\,Istituto Nazionale di Fisica Nucleare, Sezione di Napoli, 80126 Napoli, Italy}
\affil{$^{m2}$\,NASA Exoplanet Science Institute, MS 100-22, California Institute of Technology, Pasadena, CA 91125, US}
\affil{$^{m3}$\,Istituto Internazionale per gli Alti Studi Scientifici (IIASS), 84019 Vietri Sul Mare (SA), Italy}
\affil{$^{m4}$\,Max Planck Institute for Astronomy, K\"onigstuhl 17, 69117 Heidelberg, Germany}
\affil{$^{m5}$\,Yunnan Observatories, Chinese Academy of Sciences, Kunming 650011, China }
\affil{$^{m6}$\,Key Laboratory for the Structure and Evolution of Celestial Objects, Chinese Academy of Sciences, Kunming 650011, China }
\affil{$^{m7}$\,Niels Bohr Institute, University of Copenhagen, Juliane Maries Vej 30, 2100 K{\o}benhavn {\O}, Denmark }
\affil{$^{m8}$\,Centre for Star and Planet Formation, Natural History Museum, University of Copenhagen, {\O}stervoldgade 5–7, 1350 K{\o}benhavn K, Denmark  }
\affil{$^{m9}$\,Korea Astronomy and Space Science Institute, Daejeon 305-348, Republic of Korea }
\affil{$^{m10}$\,Finnish Centre for Astronomy with ESO (FINCA), University of Turku, V{\"a}is{\"a}l{\"a}ntie 20, 21500 Piikki{\"o}, Finland }
\affil{$^{m11}$\,Instituto de Astrof{\i}sica, Facultad de F{\i}sica, Pontificia Universidad Cat\'olica de Chile, Av.\ Vicu\~na Mackenna 4860, 7820436 Macul, Santiago, Chile }
\affil{$^{m12}$\,Department of Physics, Sharif University of Technology, P.\,O.\,Box 11155-9161 Tehran, Iran }
\affil{$^{m13}$\,Astrophysics Group, Keele University, Staffordshire, ST5 5BG, UK }
\affil{$^{m14}$\,Institut d'Astrophysique et de G\'eophysique, Universit\'e de Li\`ege, 4000 Li\`ege, Belgium }
\affil{$^{m15}$\,Dipartimento di Fisica ``E. Amaldi'',Universit\`a di Roma Tre, Via della Vasca Navale 84, 00149 Roma, Italy. }
\affil{$^{m16}$\, Istituto Nazionale di Fisica Nucleare, Sezione di Roma Tre, Italy. }

\begin{abstract}
For all exoplanet candidates, the reliability of a claimed detection needs to be assessed through a careful study of systematic errors in the data to minimize the false positives rate. We present a method to investigate such systematics in microlensing datasets using the microlensing event OGLE-2013-BLG-0446 as a case study. The event was observed from multiple sites around the world and its high magnification ($A_{\rm max}\sim3000$) allowed us to investigate the effects of terrestrial and annual parallax. Real-time modeling of the event while it was still ongoing suggested the presence of an extremely low-mass companion ($\sim 3 M_{\oplus}$) to the lensing star, leading to substantial follow-up coverage of the light curve. We test and compare different models for the light curve and conclude that the data do not favour the planetary interpretation when systematic errors are taken into account. 
\end{abstract}
\section{Introduction}     \label{sec:introduction}

For the past ten years, gravitational microlensing has been used to detect cool planets around G, K and M-stars in the Milky Way, allowing access to a planetary regime difficult to observe with the
transit or radial velocity methods (i.e microlensing is sensitive to planets beyond the snowline). Due to the increased field of view of the OGLE-IV and MOA-II surveys, and the recently improved performance of follow-up teams, the number of planets detected by microlensing has gone up substantially (typically 10-20 planets detected per year and 33 published to date).

Another advantage of the microlensing method is that detection of planetary companions is
possible over a larger mass-range [$\sim1M_{\oplus},\sim13M_J$], including brown dwarfs, if the projected orbital radius $s$ is in the range 0.6-1.6 $R_E$ (i.e the classical "lensing zone"). Thanks to better photometric coverage of light curves, recent studies have advanced claims about the detection of small planets \cite{Bennett2014}. However, smaller mass-ratios tend to produce smaller deviations from a single lens model most of the time. Failing to account for photometric systematics can potentially lead to false detections. The analysis of photometric systematics has been important in transit searches and has substantially improved the reliability of detections \cite{Kovacs2005,Smith2012}. This point is too often neglected by the microlensing community.

In this work, we present an extensive study of photometric systematics for the case of OGLE-2013-BLG-0446 and we compare the significance when different microlensing models are considered. Section~\ref{sec:observations} presents a summary of the observations of microlensing event OGLE-2013-BLG-0446 from multiple sites around the world. We present our modeling process in Section~\ref{sec:modeling} and conduct a study of systematics in the data in Section~\ref{sec:sysstudy}. We
present our conclusions in  Section~\ref{sec:conclusions}.


\section{Observations}    \label{sec:observations}
Microlensing event OGLE-2013-BLG-0446 ($\alpha=18^h06^m56^s.18$, $\delta
=-31^{\circ}39'27''.2$ (J2000.0); $l=0.049^{\circ}$, $b=-5.344^{\circ}$) was 
discovered on the $6^{th}$ April 2013 by the Optical Gravitational Lens Experiment (OGLE) 
Early Warning System \cite{Udalski2003b} and later alerted by the Microlensing Observations in Astrophysics (MOA) \cite{Bond2001}. Observations obtained on the rising part of the light curve indicated that this event could be highly magnified and might therefore be highly sensitive to planets \cite{Griest1998,Yee2009,Gould2010}. Follow-up 
teams, such as $\mu$FUN \cite{Gould2006}, PLANET \cite{Beaulieu2006}, RoboNet \cite{Tsapras2009} and MiNDSTEp \cite{Dominik2008}, then began observations a few days before the peak of the event. The peak magnification was $\sim$ 3000 and the peak was densely sampled from different observatories. 

The various teams used difference image analysis (DIA) to obtain 
photometry: $\mu$FUN used pySIS \cite{Albrow2009}, with the exception of the Auckland
data, which were re-reduced using {\tt DanDIA} \cite{Bramich2008,Bramich2013}. {\tt DanDIA}  
was also used to reduce the RoboNet and the Danish datasets. PLANET data were reduced online with the WISIS pipeline, and final data sets were prepared using pySIS\footnote{Data from Tasmania were obtained at the Canopus 1m observatory by John Greenhill. This was the last planetary candidate observed from Canopus before its decommissioning. These observations were also the last collected and reduced by John Greenhill (at the age 80). He has been our loyal collaborator and friend over the past 18 years and passed away on September 28, 2014.}. OGLE \cite{Udalski2015} and MOA \cite{Bond2001} used their own DIA code to reduce 
their frames. All other data sets were reduced using pySIS.

A total of 2955 data points from 16 telescopes were used for our analysis, after problematic data points were masked. A summary of each data set is available in Table~\ref{tab:sumobservations}.
\begin{table*}
  \footnotesize
  
  \centering
  \begin{tabular}{lcccccccc}
    & \\
     \hline\hline
Name&Collaboration&Location&Aperture(m)&Filter&Code&$N_{data}$&Longitude($\deg$)&Latitude($\deg$)\\
    \hline
      & \\
 $\rm{OGLE\_I}$&OGLE&Chile&1.3&I&Wo\'{z}niak&463&289.307&-29.015\\
 $\rm{OGLE\_V}$&OGLE&Chile&1.3&V&Wo\'{z}niak&24&289.307&-29.015\\
 $\rm{Canopus\_I}$&PLANET&Tasmania&1.0&I&pySIS&132&147.433&-42.848\\
 $\rm{Auckland\_R}$&$\mu$FUN&New Zealand&0.4&R&{\tt DanDIA}&107&147.777&-36.906\\
 $\rm{LSCB\_i}$&RoboNet&Chile&1.0&SDSS-i&{\tt DanDIA}&378&289.195&-30.167\\
 $\rm{LSCA\_i}$&RoboNet&Chile&1.0&SDSS-i&{\tt DanDIA}&385&289.195&-30.167\\
 $\rm{CPTA\_i}$&RoboNet&South Africa&1.0&SDSS-i&{\tt DanDIA}&22&20.810&-32.347\\
 $\rm{CTIO\_I}$&$\mu$FUN&Chile&1.3&I&pySIS&112&289.196&-30.169\\
 $\rm{CTIO\_V}$&$\mu$FUN&Chile&1.3&V&pySIS&13&289.196&-30.169\\
 $\rm{Danish\_z}$&MiNDSTEp&Chile&1.5&i+z&{\tt DanDIA}&452&289.261&-29.255\\
 $\rm{MOA\_Red}$&MOA&New Zealand&1.8&Red&Bond&454&170.464&-43.987\\
 $\rm{Possum\_N}$&$\mu$FUN&New Zealand&0.4&N&pySIS&244&177.856&-38.623\\
 $\rm{Salerno\_I}$&MiNDSTEp&Italy&0.4&I&pySIS&20&14.799&40.772\\
 $\rm{Turitea\_R}$&$\mu$FUN&New Zealand&0.4&R&pySIS&31&175.630&-40.353\\
 $\rm{Weizmann\_I}$&$\mu$FUN&Israel&0.4&I&pySIS&60&34.811&31.908\\
 $\rm{SAAO\_I}$&PLANET&South Africa&1.0&I&pySIS&58&20.789&-32.374\\
       & \\
    \hline
  \end{tabular}
  \centering
  \caption{Summary of observations. N is unfiltered dataset.}
  \label{tab:sumobservations}
\end{table*}

\section{Modeling}    \label{sec:modeling}
\subsection{Source properties}    \label{sec:sourceproperties}
This event shows clear signs of finite-source effects and the limb darkening coefficients must be evaluated for each data set. We first consider a point-source point-lens model (PSPL) 
\cite{Paczynski1986}. The PSPL model allows the estimation of the source and blended fluxes in the V and I passbands for the calibrated OGLE photometry, leading to a good approximation for the V and I magnitudes 
of the source, which in turn allows us to derive a rough color for the source. We found $(I,(V-I))_{PSPL}=(19.07,1.48)$. Using the
Interstellar Extinction Calculator on the OGLE website\footnote{http://ogle.astrouw.edu.pl/}
based on \cite{Nataf2012}, we found that the Galactic Bulge true distance modulus for this 
line of sight is $\mu=14.578\pm0.326$ mag ($d_{\rm{Bulge}}=8.2\pm1.2$ kpc), the I band extinction is 
$A_I=0.804$ mag and the reddening is $E(V-I)=0.683\pm0.036$ mag, leading to $R_I={A_I}/{E(V-I)}=1.177$, lower than the standard value of 1.5. This low extinction is known as the anomalous extinction law towards the Galactic Bulge, see \cite{Udalski2003a}. We derive the source properties as follows:
\begin{itemize}
\item Assuming that the source suffers the 
same extinction as the Red Giant Clump (i.e the source is at the same distance), we have $M_I=19.07-0.804-14.578=3.7$ mag, so the source star is most likely a main sequence star. We adopt $\log g\sim4.5$.
\item We derive its effective temperature using the dereddened color-magnitude relation for dwarfs and subgiants (relation (3) in \cite{Casagrande2010}) with solar metallicity.
\item From \cite{Claret2000} and using $\log g\sim4.5$, we are able
to find the linear limb-darkening coefficients $u_{\lambda}$ \cite{Milne1921}
for each filter. Following \cite{Albrow1999,Yoo2004}, we use the transformation :
\begin{equation}
\Gamma_{\lambda}={{2u_{\lambda}}\over{3-u_{\lambda}}}
\end{equation}
\end{itemize}

These calculations form the starting point for an iterative fit of the FSPL (finite source point lens) model, together with error-bar rescaling as described in Section~\ref{sec:rescaling}. Our best FSPL 
model converges to source magnitude and color $(I,(V-I))=(19.00,1.49)$. Correcting for extinction and reddening we have
 $(I_o,(V-I)_o)=(18.20,0.81)$. The corresponding effective temperature of the source is $T_{\rm{eff}}\sim5400~K$, 
leading to $\Gamma_{V}=0.63~(u_{V}=0.72)$, $\Gamma_{R}=0.55~(u_{R}=0.65)$ 
and $\Gamma_{I}=0.46~(u_{I}=0.56)$ for $\log g\sim4.5$. Note that we also use $\Gamma_{I}=0.46$ for the RoboNet telescopes (SDSS-i filter).
Finally, given the dereddened magnitude and colour of the source from our 
best FSPL model, we are able to estimate the angular source star radius $\theta_*$ using \cite{Kervella2008}:
\begin{equation}
\log_{10}(\theta_*)=3.1982+0.4895(V-I)_o-0.0657(V-I)_o^2-0.2I_o
\end{equation}

The uncertainty of this relation is 0.0238. The errors on our magnitude estimates are $(\Delta I,\Delta V)=(0.02,0.02)$ mag. 
Assuming a conservative estimate of the 
error on $A_I$ (0.1 mag) and using standard error propagation gives 9\% precision: $\theta_*=0.82\pm0.07~\mu as$. With
the adopted source distance (8.234 kpc), the source star radius is $R_*=1.4\pm0.3~R_\odot$. Therefore the source is a G6 or K0 star \cite{Bessell1988}.    
\subsection{Single lens model}    \label{sec:fspl}
\begin{figure*}[ht]
  \centering
  \includegraphics[width=19cm]{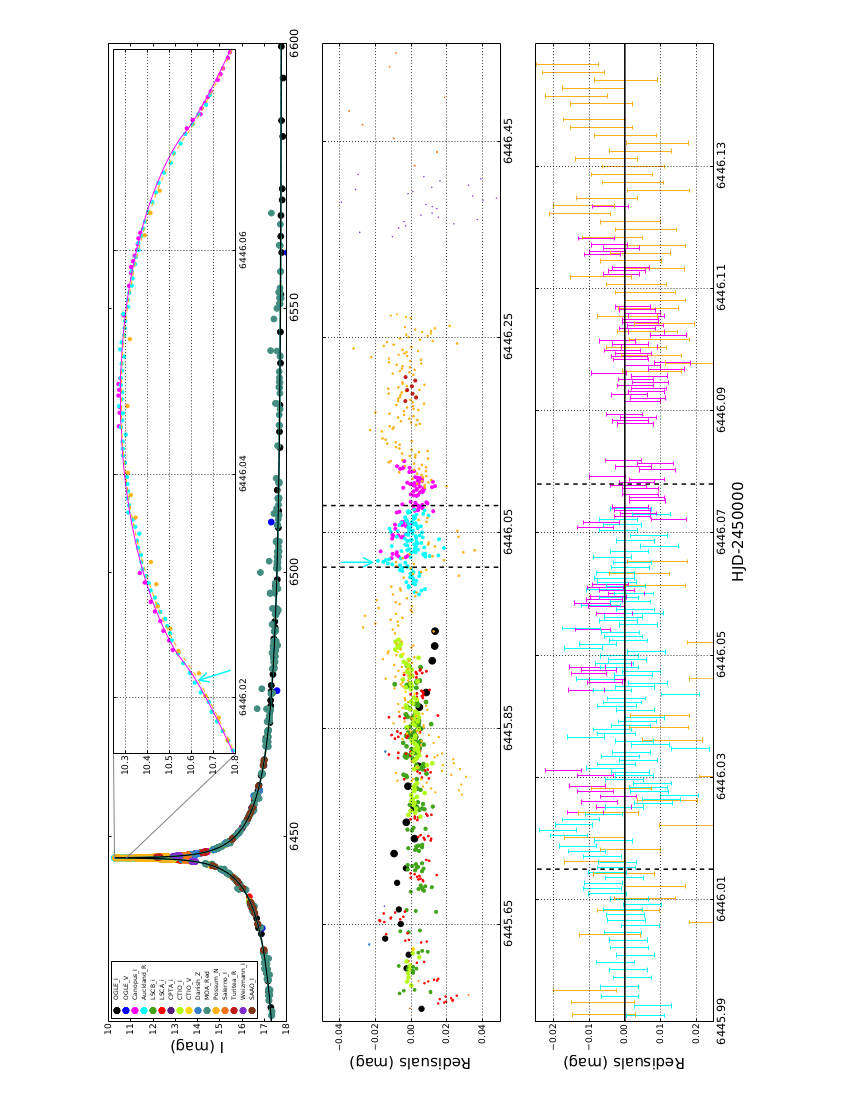}
    \caption{Light curve of OGLE-2013-BLG-0446 with our best FSPL model. The top panel shows the full 2013 light curve with a maximum magnification at $HJD-2450000\sim6446.0$ d.
The insert on the right is a zoom of the peak. The pink model light curve is for $\Gamma_{I}=0.46$ and the orange dashed model light curve is for $\Gamma_{V}=0.63$.  The cyan model for the R band is not shown for clarity. The cyan arrow indicates the position of the possible planetary anomaly. The middle panel shows residuals of the FSPL model close to the peak. The radius of each point is proportional to the inverse square of the error bar (bigger points have smaller error bars). The bottom panel is a closer view of the possible anomaly. For the bottom two panels, vertical dashed black lines indicate the time window corresponding to the insert in the top figure. }
    \label{fig:lightcurve}
\end{figure*}

The PSPL model is described by the standard single-lens parameters : $t_E$ the Einstein 
crossing time, $u_o$ the minimum impact parameter and $t_o$, the time of this
minimum. The normalized angular source radius $\rho={\theta_*}/{\theta_E}$ \cite{
Nemiroff1994,Witt1994,Gould1994,Bennett1996,Vermaak2000}, where $\theta_E$ is the 
angular Einstein ring radius, is included in the model along with the previous parameters to take into account finite-source effects close to the magnification peak. We used the method 
described in \cite{Yoo2004} to take into account the change in magnification due to the extended source. The FSPL model significantly improves the fit (see Table~\ref{tab:models}). The best FSPL model is shown in Figure~\ref{fig:lightcurve}.

Using the value of $\rho$ from the FSPL model, we are able to estimate the angular radius of the Einstein ring $\theta_E={\theta_*}/{\rho}=1.57\pm0.1$ mas and the lens-source proper motion $\mu={\theta_E}/{t_E}=7.4\pm0.7~\rm{mas~yr}^{-1}$.
\subsection{Treatment of photometric uncertainties and rejection of outliers}    \label{sec:rescaling}
Because of the diversity of observatories and reduction pipelines used in
microlensing, photometric uncertainties need careful rescaling to accurately represent the real
dispersion of each data set. This is an important preliminary step in modeling the event. Following  \cite{Bachelet2012a,Miyake2012,Yee2013}, we rescale the uncertainties using :
\begin{equation}
e'_i=\sqrt{(fe_i)^2+e_{min}^2}
\end{equation}
where $e_i$ are the original magnitude uncertainties, $f$ is the rescaling parameter for low magnification 
levels, $e_{min}$ is a minimal uncertainty to reproduce the practical limitations of photometry 
and $e'_i$ are the adjusted magnitude uncertainties. The classical rescaling method is to adjust
$f$ and $e_{min}$ to force ${\chi^2}/{dof}$ to be unity. 

In this paper, we follow an alternative method of first adjusting $f$ and $e_{min}$
to force the residuals, normalized by $e'_i$, to follow a Gaussian distribution 
around the model. If possible, we also aim to obtain a ${\chi^2}/{dof}\sim1$.
Note that these two methods lead to the same results, except for the OGLE\_I dataset.
For OGLE\_I, the distribution without rescaling shows some data points with large residuals. This is not surprising because the OGLE\_I data set covers the entire light curve with a large number of points,  especially the faint baseline magnitude ($I\sim17.8$), with a constant exposure time in order of 100 s. Inspection of the OGLE\_I light curve reveals that the uncertainties during high 
magnification are underestimated, so we adjust the $e_{min}$ parameter. We tried
to force ${\chi^2}/{dof}\sim1$ for this dataset, but this generated
large uncertainties for the low magnification part (i.e. the baseline) leading to a non-Gaussian distribution (lots of normalised residuals too close to the mean).

We finally checked isolated points far away from this Gaussian distribution, and reject as outliers ($> 7 \sigma $) two data points in the Auckland\_R dataset.
The rescaling coefficients are presented in Table~\ref{tab:rescaling}.
\begin{table}
  \footnotesize
  
  \centering
  \begin{tabular}{lcccccc}
    & \\
     \hline\hline
Name&$N_{data}$&$\Gamma_{\lambda}$&f&$e_{min}$\\
    \hline
      & \\
 $\rm{OGLE\_I}$&463&0.46&1.0&0.002\\
 $\rm{OGLE\_V}$&24&0.63&10.25&0.0\\
 $\rm{Canopus\_I}$&132&0.46&3.0&0.005\\
 $\rm{Auckland\_R}$&107&0.55&1.75&0.005\\
 $\rm{LSCB\_i}$&378&0.46&1.4&0.003\\
 $\rm{LSCA\_i}$&385&0.46&2.0&0.007\\
 $\rm{CPTA\_i}$&22&0.46&1.19&0.0\\
 $\rm{CTIO\_I}$&112&0.46&1.5&0.004\\
 $\rm{CTIO\_V}$&13&0.63&1.0&0.0\\
 $\rm{Danish\_z}$&452&0.46\footnote{The transmission curve for this filter is close to a Johnson Cousin I, see \cite{Skottfelt2015}.}&5.0&0.008\\
 $\rm{MOA\_Red}$&454&0.51\footnote{We select a bandpass between I and V.}&1.0&0.0\\
 $\rm{Possum\_N}$&244&0.63\footnote{For this unfiltered data, we choose the filter closest to the CCD spectral response.}&1.5&0.008\\
 $\rm{Salerno\_I}$&20&0.46&3.91&0.0\\
 $\rm{Turitea\_R}$&31&0.55&1.0&0.005\\
 $\rm{Weizmann\_I}$&60&0.46&3.2&0.01\\
 $\rm{SAAO\_I}$&58&0.46&2.57&0.008\\
       & \\
    \hline
  \end{tabular}
  \centering
  \caption{ Limg darkening and error bar rescaling coefficients used in this paper.}
  \label{tab:rescaling}
\end{table}

\subsection{Annual and terrestrial parallaxes}    \label{sec:parallax}
We looked for second-order effects in the light curve. First, the relatively
long Einstein-ring crossing-time ($t_E\sim80$ days) should allow the measurement of the
displacement of the line-of-sight towards the target due to the Earth's rotation 
around the Sun. This annual parallax \cite{Gould1992,Gould2000,Smith2003,Gould2004,Skowron2011} is described by the vector $\pi_{E,a}={AU}/{\widetilde{r_E}}=(\pi_{EN},\pi_{EE})$,
where $\widetilde{r_E}$ is the angular radius of the Einstein ring of the lens projected onto
the observer plane and $\pi_{EN}$ and $\pi_{EE}$ are the components of this 
vector in the North and East directions respectively. In practice, the introduction of this parameter slightly changes the value of the impact parameter and $\tau={(t-t_o)}/{t_E}$.
Strong modifications of the light curve can be seen far from the peak of the 
event, i.e in the wings of the light curve, with few changes around the peak, 
see for example \cite{Smith2003}. To model this effect, the constant $t_{o,par}$ 
\cite{Skowron2011} is added to give an invariant reference time for each model.
We choose $t_{o,par}=2456446.0$ HJD for our models.

Since this event is so highly magnified, it should also be possible to
measure the terrestrial parallax. \cite{Hardy1995} first introduced the idea 
that for an 'Extreme Microlensing Event' (EME), the difference in longitudes of
observatories should result in light curves where tiny changes in the line-of-sight towards the target become apparent, allowing a measurement of the Einstein ring \cite{Holz1996,Gould1997,Dong2007}. Again, 
this effect is described by the parallax vector $\pi_{E,t}={AU}/{r_\oplus}({\Delta t_o}/{t_E},\Delta U_o)$, 
where $r_\oplus$ is the Earth radius. 
\cite{Gould2013} estimated that the condition $\rho\widetilde{r_E}\leq50r_{\oplus}$ 
is required to expect a measurable difference in terms of magnification. This 
condition leads to $\pi_E>0.24$ for this event by using an approximate 
value for the normalized source star radius $\rho\sim5\times10^{-4}$. A summary of longitudes and latitudes of the observatories is in Table~\ref{tab:sumobservations} and results
are summarized in Table~\ref{tab:models}. 

Note that we also compute the annual parallax model for a positive impact parameter ($u_0>0$) and found no significant difference with the model reported in the Table~\ref{tab:models}. This is the $u_0$ degeneracy described in the literature \cite{Smith2003,Gould2004,Skowron2011}. For the terrestrial parallax, a positive impact parameter leads to a better fit ($\Delta\chi^2\sim60$) for equivalent $\pi_{EN}$ and $\pi_{EE}$ values , which is a similar result to \cite{Yee2009}. 
\subsection{Binary model}    \label{sec:binary}
\begin{figure}[ht]
  \centering
  \includegraphics[height=7cm]{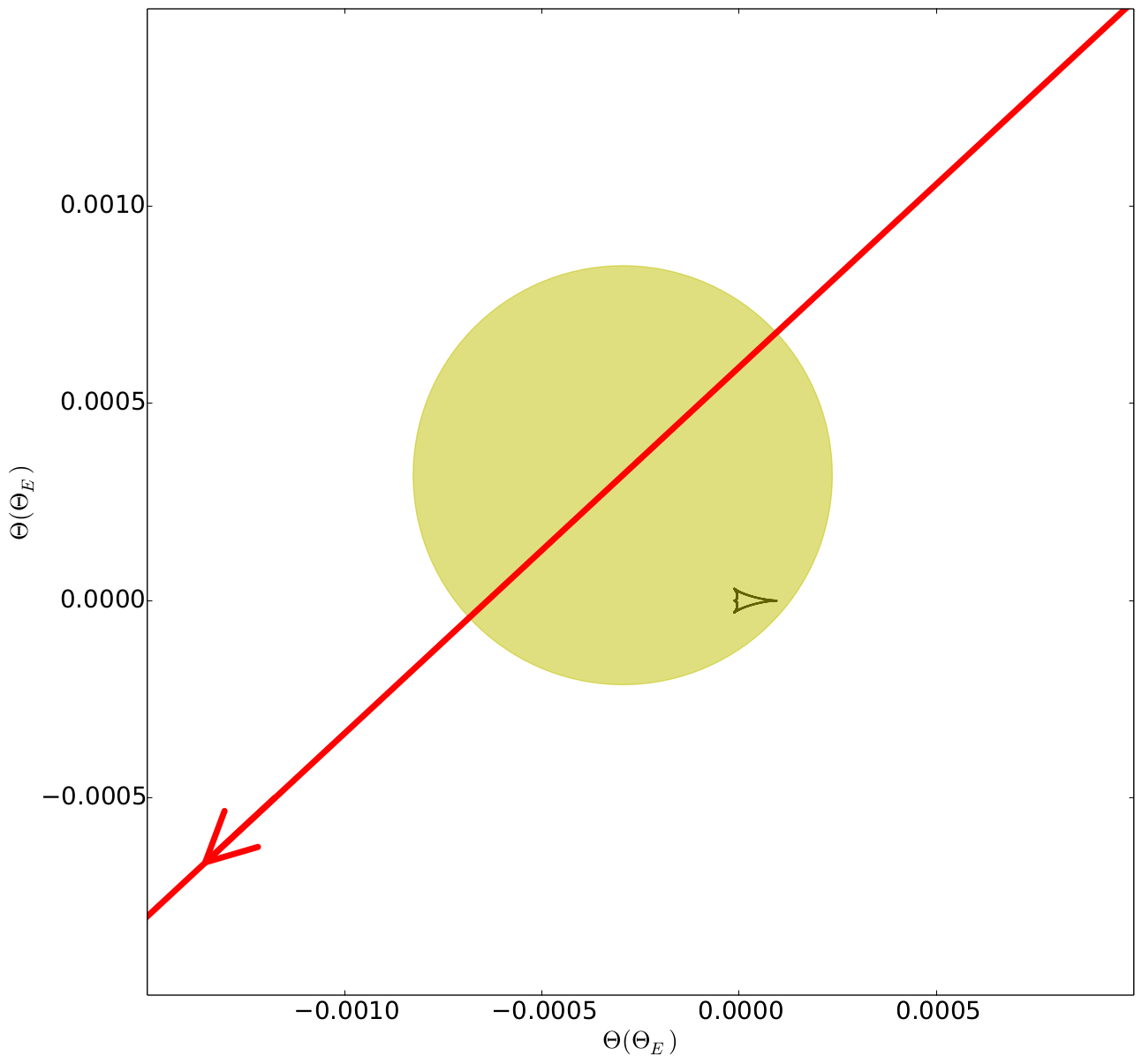}
    \caption{Lens geometry for our best fit planetary model. The yellow disc represents the source, the red line indicates the source trajectory and the black closed curve represents the central caustic. The caustic signature in the light curve is highly 'diluted' by the relatively large source star.}
    \label{fig:lensgeometry}
\end{figure}

\begin{figure*}[ht]
  \includegraphics[width=\textwidth]{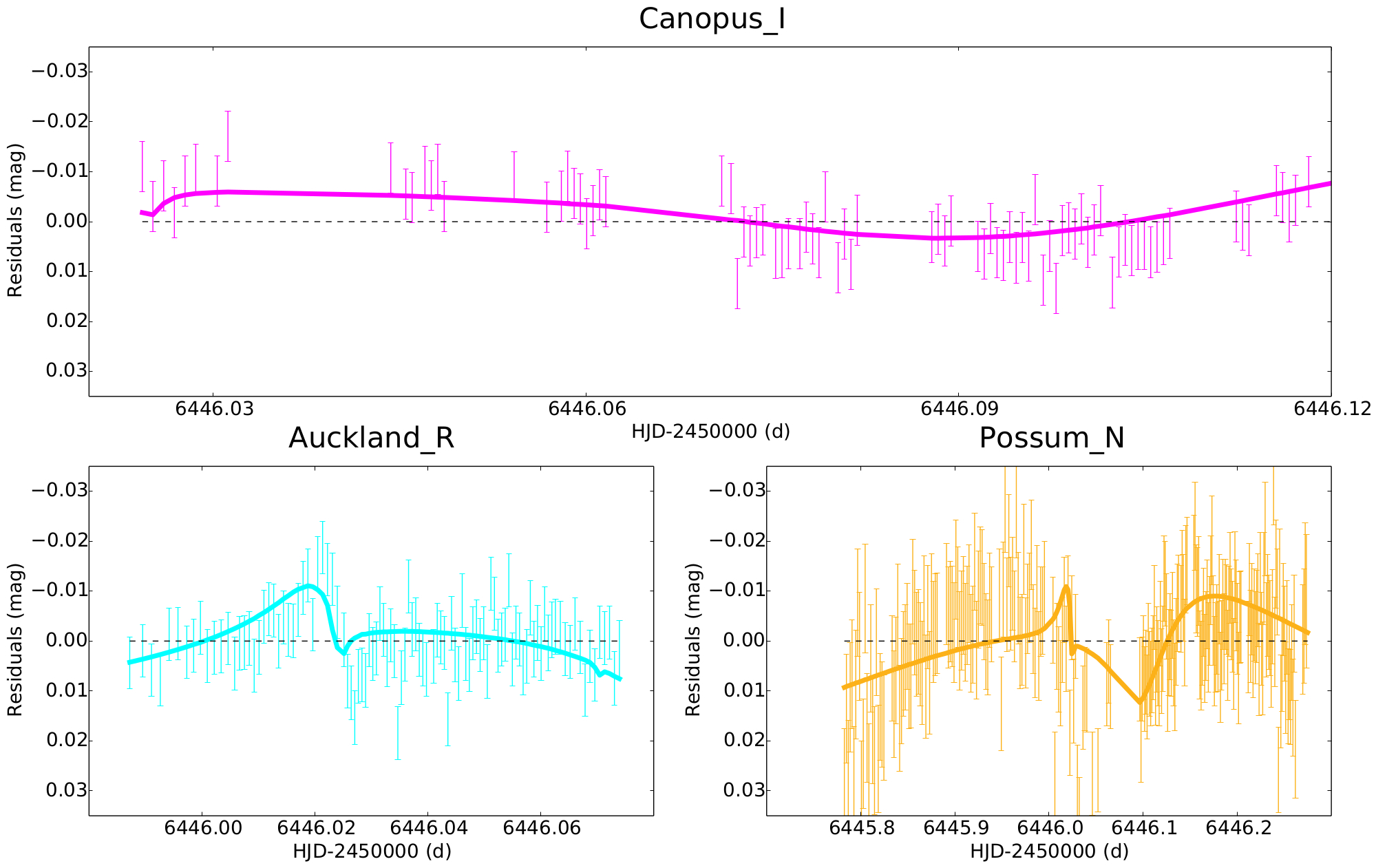}
    \caption{FSPL residuals close to the peak. The curves represent the best planetary model.}
    \label{fig:residuals}
\end{figure*}

 At the end of the 2013 observing season, several planetary models circulated (private communication) indicating the presence of the
smallest microlensing planet ever detected ($q \sim 2 \times 10^{-5}$). In order to investigate these claims of the existence of a very low-mass ratio  planetary companion to the primary lens, and to exclude the possibility of a false alarm, we used a 
finite-source binary lens model (FSBL) with three extra 
parameters: the projected separation (normalized by $\theta_E$) between the 
two bodies $s$ , the mass ratio $q$ using the convention described in \cite{Bachelet2012b} 
(the most massive component on the left) and $\alpha$ the source trajectory angle 
measured from the line joining the two components (counterclockwise angle). We first used a grid search and we
finally explore minima with a full Markov-Chain Monte Carlo algorithm (MCMC).
Please see for example \cite{Dong2009b} or \cite{Bachelet2012b} for more details.
We find two local minima which correspond to the known theoretical degeneracy $s\Leftrightarrow s^{-1}$. We only explore the 'wide' solution, which gives the best grid-search $\chi^2$, for three reasons. First, as explained
in the next section, the reliability of the planetary model is not clear. Second, we expect a strong degeneracy in terms of $s\Leftrightarrow s^{-1}$, so models should converge to solutions with similar shapes for the central caustic and therefore similar residuals. Finally, due to the really
small value of $\rho$, modeling this event is very time consuming.
We present our results in Table~\ref{tab:models}, our best caustic-crossing geometry in Figure~\ref{fig:lensgeometry} and redisuals to the FSPL model for data sets covering the magnification peak are plotted in  Figure~~\ref{fig:residuals}. 

\begin{table*}
  
  \centering
  \footnotesize
  \begin{tabular}{lcccc}
    & \\
     \hline\hline
     Parameters&FSPL&FSPL+Annual parallax&FSPL+Terrestrial parallax&Wide planetary (FSBL)\\
     \hline
      & \\
      $t_o$(HJD)&$6446.04790\pm3~10^{-5}$&$6446.04678\pm3~10^{-5}$&$6446.04681\pm3~10^{-5}$&   $6446.04659\pm3~10^{-5}$\\
      $U_o$($\theta_E$)&$-4.21~10^{-4}\pm7~10^{-6}$&$-4.02~10^{-4}\pm8~10^{-6}$&$-4.22~10^{-4}\pm8~10^{-6}$&
$-4.31~10^{-4}\pm5~10^{-6}$\\
     $t_E$(days)&$76.9\pm1.3$&$80.4\pm1.5$&$76.5\pm1.4$&$76.0\pm0.7$\\
     $\rho$($\theta_E$)&$5.22~10^{-4}\pm9~10^{-6}$&$4.99~10^{-4}\pm1~10^{-5}$&$5.24~10^{-4}\pm9~10^{-6}$&$5.31~10^{-4}\pm5~10^{-6}$\\
     $I_s$(mag)&$19.00\pm0.02$&$19.05\pm0.02$&$19.00\pm0.02$&$18.99\pm0.01$\\
     $V_s$(mag)&$20.49\pm0.02$&$20.54\pm0.02$&$20.49\pm0.02$&$20.48\pm0.01$\\
     $I_b$(mag)&$18.21\pm0.01$&$18.18\pm0.01$&$18.21\pm0.01$&$18.22\pm0.01$\\
     $V_b$(mag)&$22.72\pm0.11$&$22.35\pm0.11$&$22.76\pm0.13$&$22.85\pm0.07$\\
     $\Pi_{EN}$&&$0.37\pm0.15$&$0.07\pm0.02$&\\
     $\Pi_{EE}$&&$0.27\pm0.005$&$0.01\pm0.02$&\\
     $s$($\theta_E$)&&&&$1.68\pm0.05$\\
     $q$&&&&$3.1~10^{-5}\pm2~10^{-6}$\\
     $\alpha$(rad)&&&&$-2.39\pm0.02$\\
     $\chi^2$&3900.781&3839.267&3877.241&3551.217\\
       & \\
    \hline
  \end{tabular}
  \centering
  \caption{Model parameters.}
  \label{tab:models}
\end{table*}

\section{Study of systematic trends in the photometry} \label{sec:sysstudy}
\subsection{Generality and method} \label{sec:sysgeneral}
Our best planetary model claims the detection of smooth deviations in the light curve away from the FSPL model at a peak-to-peak level of $\le1\%$,  which is supposedly caused by the source
passing over the central planetary caustic. It is well known to photometrists, however, that from ground-based telescopes the photometric precision at this level can be affected by systematic trends (or red noise) in the data.

In the early days of planet hunting using the transit method, researchers were confounded as to why they were not finding as many planets as predicted. The predictions were of course based on simulated light
curves taking into account stochastic noise from the photons (sky and star) and the CCD, but ignoring the effects of sub-optimal data calibration/reduction that introduce correlated noise
(e.g. \cite{Mallen2003} and \cite{Pepper2005}).
It was soon realised that transit detection thresholds were severely affected by systematic trends in the light curves \cite{Pont2006,Aigrain2007} with the knock-on effect of reducing the predicted planetary yield of a transit survey, and at least partially explaining the unexpectedly low rate of transiting planet discoveries. The microlensing
planet hunters face a similar problem for detecting low-amplitude ($\la$1\%) planetary deviations in microlensing light curves, especially when no ``sharp'' light curve features, caused by caustic
crossing events, are predicted/observed. However, the microlensing community is now aiming for really low amplitude signal detection which requires extra care in the treatment of systematic errors \cite{Yee2013}.

Systematic trends in light curve data can be caused by an imperfect calibration of the raw data and sub-optimal extraction of the photometry. For instance, on the calibration side,
flat fielding errors which vary as a function of detector coordinates can induce correlated errors in the photometry as the telescope pointing drifts slightly during a set of time-series exposures.
On the software side, systematic errors in the photometry can be caused by errors in the PSF model used during PSF fitting for example. Also the airmass and transparency variations in the datasets should be modeled in the DIA procedure by the photometric scale factor. However, there is no garantee that the DIA modeling is perfect and this can create systematics trends in the data. As recently discussed by \cite{Bramich2015}, an error $\epsilon_p$ in the estimate of the photometric scale factor leads directly to an error $\epsilon_p$ in the photometry. For example, the passage of clouds during  data acquisition can create inhomogeneous atmospheric transparency in the frames and lead to a spatially varying photometric scale factor. The estimation of the photometric scale factor in DIA by using a 'mean' value for the whole frame will produce different systematic trends for each star in the field of view. In practice, the expected error $\epsilon_p$ is order of a few \%, which is non-critical for the majority of microlensing deviations, but can easily imitate the smallest such as in OGLE-2013-BLG-0446.

Obtaining a photon-noise limited data calibration and photometric extraction
is not always feasible. Therefore complementary techniques have been developed to perform a relative calibration of the ensemble photometry after the data reduction (i.e. a post-calibration). These techniques
can be divided into two broad groups; namely, detrending methods that do not use any {\it a priori} knowledge about the data acquisition or instrumental set up (e.g. \cite{Tamuz2005}), and photometric modelling
methods that attempt to model the systematic trends based on the survey/instrumental properties (e.g. \cite{Honeycutt1992,Padmanabhan2008,Regnault2009}). Each data point is associated with a unique object and a unique image (epoch), and carries associated metadata such as magnitude uncertainty, airmass, $(x,y)$ detector
coordinates, PSF FWHM, etc. To investigate the systematic trends in the photometry, we firstly identified a set of object/image properties which we suspected of having influenced
the quality of the data reduction. For each of these quantities, we defined a binning that covers the full range of values with an appropriate bin size. For each bin, we introduced an
unknown magnitude offset to be determined, the purpose of which is to model the mean difference of the photometric measurements within the corresponding bin from the rest of the 
photometric measurements. We constructed our photometric model by adopting the unknown true instrumental magnitude of each object\footnote{Except for one object where we fixed the true instrumental
magnitude to an arbitrary value to avoid degeneracy.} and the magnitude offsets as parameters. Since the model
is linear, the best-fit parameter values corresponding to the minimum in $\chi^{2}$ may be solved for directly (and in a single step using some matrix algebra - see Bramich \& Freudling 2012).
Iteration is of course mandatory to remove variable stars and strong outliers from the photometric data set. A valid criticism of this method is that the systematic trends are derived from the constant stars but then applied to all stars including the variable stars (the microlensing event in our case). The question arises as to whether this approach is consistent? To argue our case, we are limited to showing that the method works in practice and we direct the reader to Figure 1 of \cite{Kains2015} where RR Lyrae light curves in M68 are much improved by this self-calibration method.
We used this method to analyse systematic trends in three data sets. They are listed below.

\subsection{LSCA\_i and LSCB\_i : the twins paradox} \label{sec:sysgeneral}
We opted to employ the above methodology in order to investigate and understand the systematic trends in the LSCB\_i and LSCA\_i data sets using the algorithms described in \cite{Bramich2012}\footnote{The code is a part of DanIDL, available at http://www.danidl.co.uk/}. These telescopes are twins : both are LCOGT 1m telescope clones, both supporting Kodak SBIG STX-16803 CCDs at the time of these observations. SDSS-i prescription filters manufactured at the same time were use to observe OGLE-2013-BLG-0446 during the same period of observation, though not precisely synchronously. 

\begin{figure}[ht]
  \includegraphics[height=6cm]{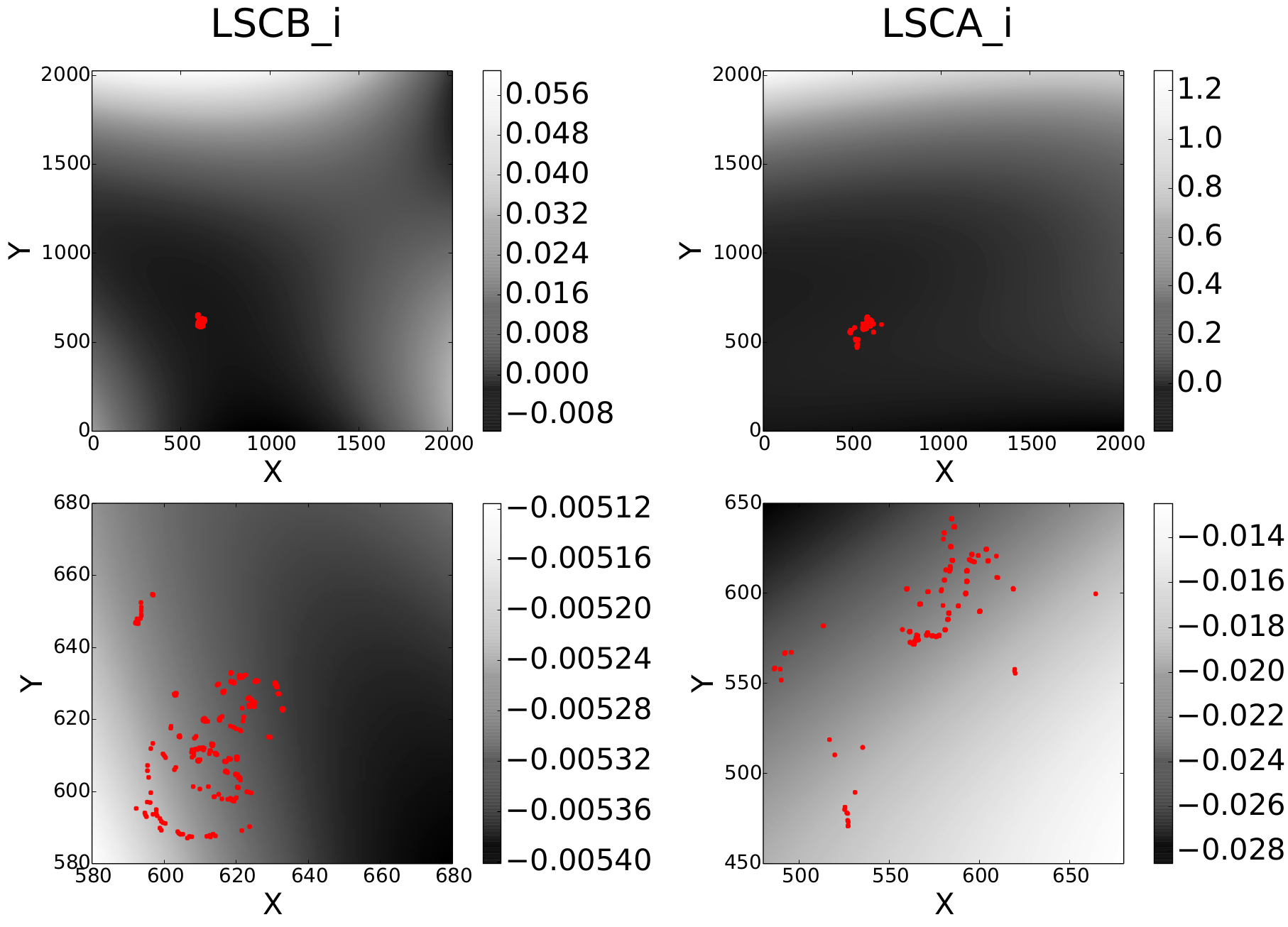}
    \caption{Illumination correction for LSCB\_i and LSCA\_i data sets. The bottom figures are zoom close to the pointing area.}
    \label{fig:LSCill}
\end{figure}

We first chose to study  LSCB\_i because this telescope most strongly favors the planetary model ($\Delta\chi^2\sim128.3$, see Table~\ref{tab:chi2corr}). For LSCB\_i, the {\tt DanDIA} pipeline extracted 4272 light curves from the images in the LSCB\_i data set, each with 378 data points (or epochs), which yields a total of
1614816 photometric data points.
We investigated each object/image property in turn using the above method, and determined the peak-to-peak amplitude of the magnitude offsets in each case. The results are reported in Table~~\ref{tab:trends}. The trends in the photometry were found to be at the sub-mmag level for all correlating properties except for the epoch (2.0~mmag). The magnitude offsets
determined for each epoch (or image) serve to correct for any errors in the fitted values of the photometric scale factors during DIA. The magnitude offsets as a function of detector
coordinates (commonly referred to as an illumination correction - e.g. \cite{Coccato2014}) were modelled using a two-dimensional cubic surface
(as opposed to the binning previously described) so as to better capture the large-scale errors in the flat-fielding. The peak-to-peak amplitude of the cubic surface over the full detector
area was found to be $\sim$60~mmag, but since the LSCB\_i observations only drifted by $\sim$50 pixels in each coordinate, we found that the magnitude offsets applicable to the OGLE-2013-BLG-446
light curve have a peak-to-peak amplitude of only $\sim$0.2~mmag. This can be seen in Figure~~\ref{fig:LSCill}. The overall level of systematic trends in the LSCB\_i data set
for OGLE-2013-BLG-446 is $\sim$2.0mmag. To conclude, this analysis reveals that the illumination correction is not sufficient to explain the observed systematics.

\begin{table*}
  \footnotesize

  \centering
  \begin{tabular}{lcccccc}
    & \\
     \hline\hline
Correlating Quantity         & Possible Underlying Cause                           & \multicolumn{3}{c}{Peak-To-Peak Amplitude(mmag)}       \\
                             &                                                     & LSCA\_i & LSCB\_i & Auckland\_R \\
    \hline  
Exposure time                & CCD non-linearities                            & 20 & 0.3 & 60  \\
Airmass                      & Varying extinction                             & 22 & 0.8 & 40 \\
PSF FWHM                     & Varying seeing disk                            & 25 & 0.4 & 28 \\
Photometric scale factor     & Reduction quality at different transparencies  & 20 & 0.2 & 40 \\
Epoch                        & Errors in photometric scale factor             & 60 & 2.0 & 120 \\
Detector coordinates         & Flat-field errors                              & 10 & 0.2 & * \\
Background		     & Reduction quality                              & 27 & 0.5  & 45 \\
    \hline  
  \end{tabular} 
 \caption{The peak-to-peak amplitude of the magnitude offsets for each object/image property that we investigated for causing systematic trends in the photometry for the LSCB\_i, LSCA\_i and Auckland\_R. data sets We also list a possible underlying cause for any systematic trends that are found as a function of the corresponding object/image property.}
  \label{tab:trends}
\end{table*}

We chose also to study the LSCA\_i data set because this telescope observed the target at the same time but does not show any planetary significance ($\Delta\chi^2\sim20.7$). We conduct the same study and the results are summarized in Table~~\ref{tab:trends} and Figure~~\ref{fig:LSCill}. The peak-to-peak amplitudes of the magnitude offsets are ten times bigger than for the LSCB\_i dataset. It is surprising to see how two similar intruments can lead to such different data quality. A more careful check of the frames clearly shows a problem in the focus of the LSCA\_i telescope. Because the Galactic Bulge fields are very crowded, this is a critical point for microlensing observations (e.g. increasing the blending). A plausible explanation of this difference between the twins is that during the time of observations, the telescopes were under commissioning, leading to non-optimal performance for LSCA\_i.

\subsection{Auckland\_R} \label{sec:sysgeneral}
\begin{figure*}
  \includegraphics[width=\textwidth]{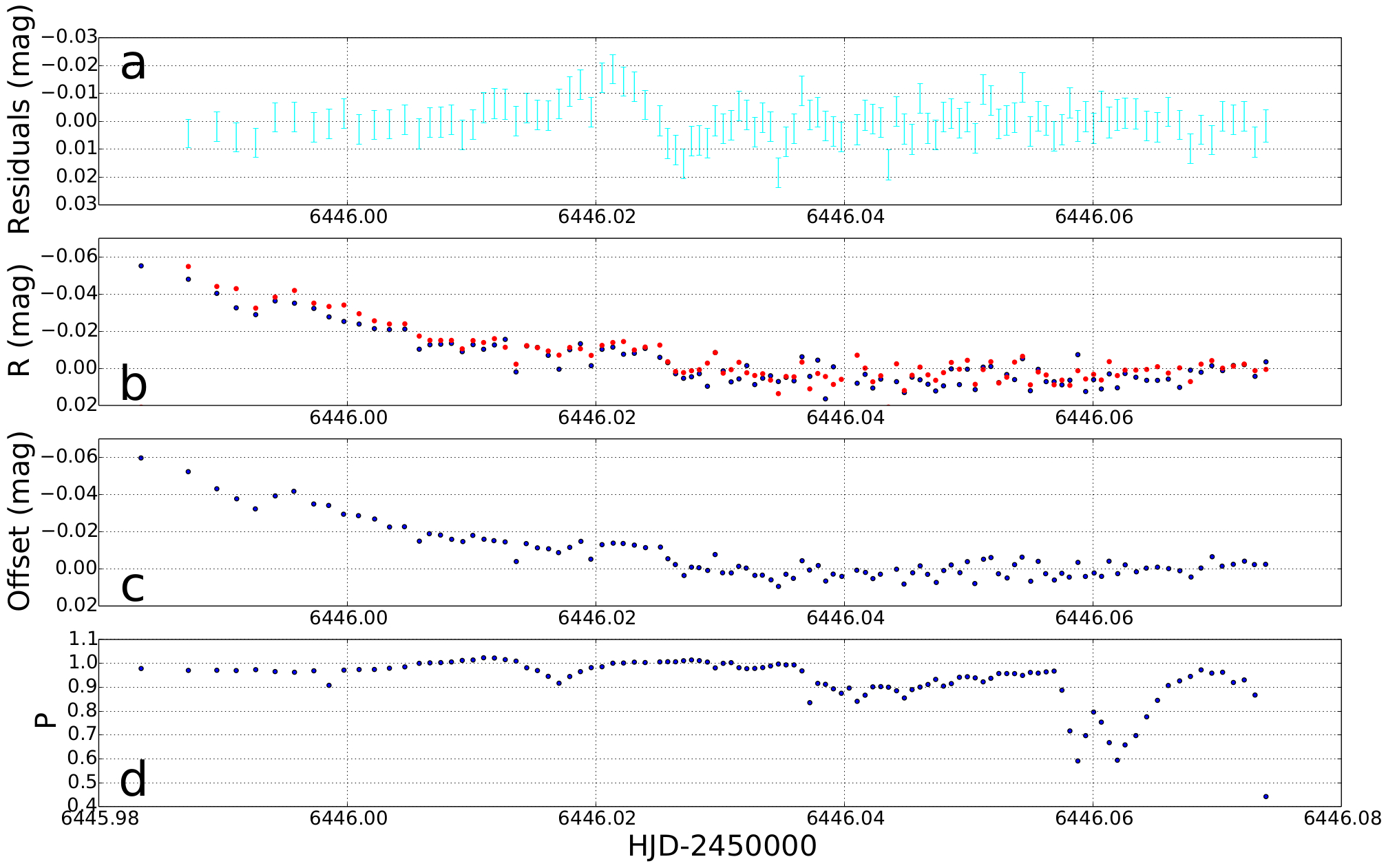}
    \caption{a: Residuals of the FSPL model for the Auckland\_R dataset. b: Light curves of the two other brightest stars in the field. c: Systematics magnitude offsets as a function of the epoch computed for this dataset, see text. d: Photometric scale factor (normalised to a single exposure). }
    \label{fig:Auckland_sys}
\end{figure*}

We conducted the same study for the Auckland\_R dataset because this telescope presents the clearest feature that mimics a planetary deviation, around $HJD\sim6446.02$ as can be seen in Figure~~\ref{fig:residuals}. Results can be seen in Table~~\ref{tab:trends}. Because the pointing
for this dataset was extremely accurate (offset less than 2 pixels for the whole night of observation), the estimation of the illumination correction was not possible. There is not enough information in the matrix equations and they are degenerate. But this reflects the fact that the pointing did not induce systematic trends.
However,  a clear variation in the magnitude offset at each epoch is visible at the time of the deviation. Furthermore, we find that this offset is stronger for the brighter stars, as can be seen in Figure~~\ref{fig:Auckland_sys}. There are strong similarities between the FSPL residuals and the magnitudes of the two brightest stars around the time of the anomaly $HJD\sim6446.02$, especially when the FSPL residuals get brighter at $HJD\sim6446.03$. Because the microlensing target is by far the brightest object in our field, we can expect that this systematics effect is probably even larger in our target. For this dataset, we slightly modified our strategy by computing the offset at each epoch only for the brightest stars ($mag<18$) and we rejected the microlensing target from the computation. Also, as can be seen in the bottom panel of Figure~~\ref{fig:Auckland_sys}, the photometric scale factor shows variations during the night. This indicates the passage of clouds which can lead to systematics errors, as described previously. For example, the FSPL residuals in the interval $6446.01\le HJD \le 6446.02$ clearly share the pattern with the photometric scale factor.

\subsection{Correction of systematics} \label{sec:sysgeneral}
\begin{figure*}[ht]
  \centering
  \includegraphics[width=10cm]{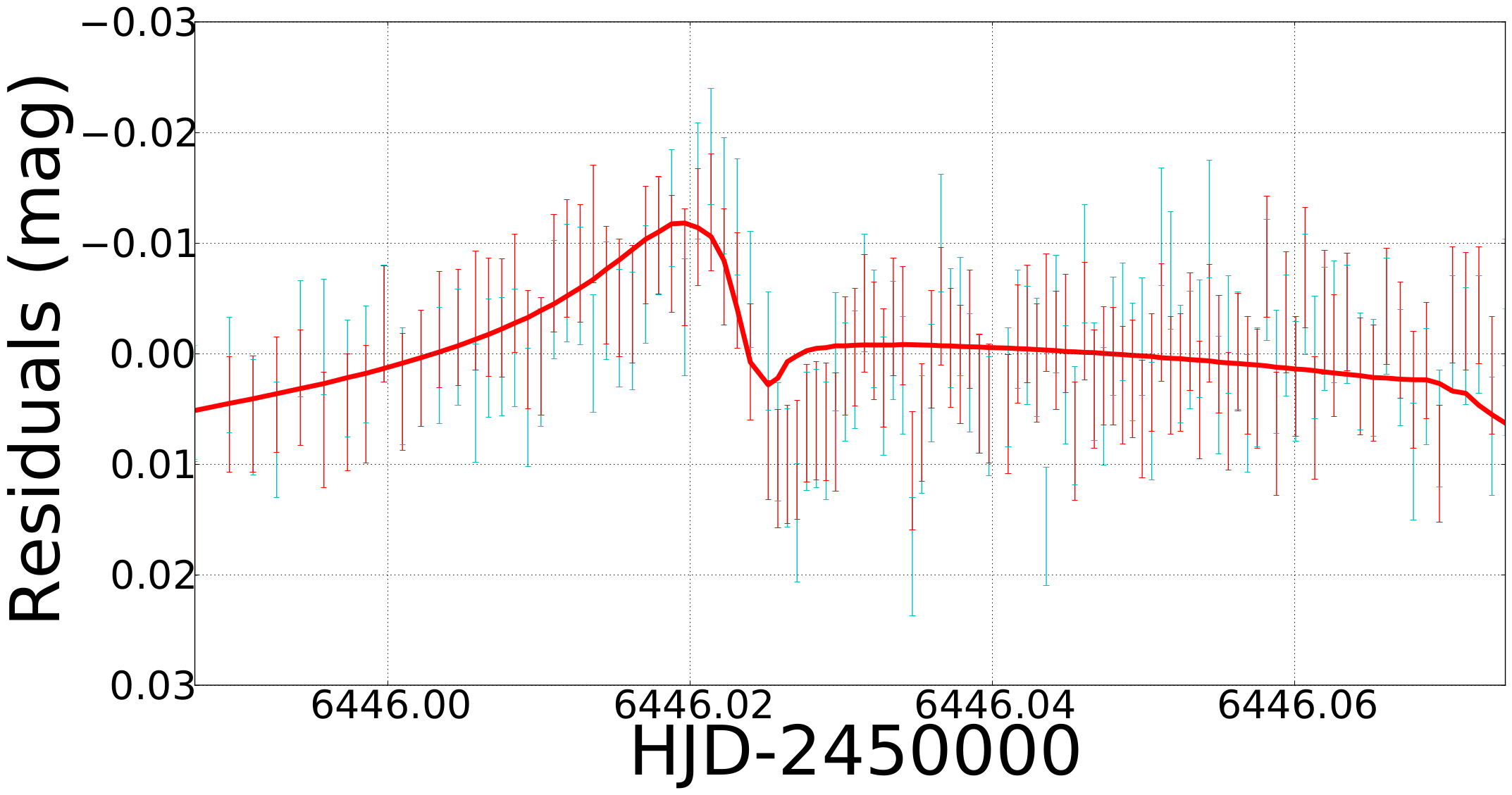}
    \caption{FSPL residuals of the Auckland\_R before (cyan) and after (red) correction for systematics using the magnitude offsets as a function of epoch. The best planetary model after correction (FSBL\_c) is shown in red.}
    \label{fig:Auckland_corr}
\end{figure*}

For the three studied datasets (LSCA\_i, LSCB\_i and Auckland\_R), we corrected the systematics for the quantity that yielded the largest peak-to-peak amplitude in the magnitude offsets: namely the epoch. Moreover, this quantity is correlated with other quantities (airmass for example) and so the epoch correction should decrease the systematic trends measured for the other paramters listed in Table~~\ref{tab:trends}. We checked this and found that for the three datasets, the epoch correction leads to a significant improvement (order of a factor ten) for the systematics of the correlated quantities. This is a first order correction and we wanted to see the impact on the different models we analyzed. We repeated our modeling process with these new datasets using the previous models as starting points. The results are presented in Table~\ref{tab:modelscorr}. The new Auckland\_R residuals can be seen in Figure~~\ref{fig:Auckland_corr}. After correction, the amplitude of the 'anomaly' is smaller but still exists. This is probably due to the fact that the amplitude of this feature in the light curves is brightness dependent , and the microlensing target is much brighter than all of the other stars, leading to an insufficient correcxtion fot the microlensing event.
\begin{table*}
  
  \centering
  \footnotesize
  \begin{tabular}{lcccc}
    & \\
     \hline\hline
     Parameters&FSPL\_c&FSPL\_c+Annual parallax&FSPL\_c+Terrestrial parallax&Wide planetary (FSBL\_c)\\
     \hline
      & \\
      $t_o$(HJD)&$6446.04818\pm2~10^{-5}$&$6446.04681\pm3~10^{-5}$&$6446.04680\pm3~10^{-5}$&   $6446.04665\pm3~10^{-5}$\\
      ${\bf u_o}$($\theta_E$)&$-4.21~10^{-4}\pm1~10^{-6}$&$-4.01~10^{-4}\pm8~10^{-6}$&$-4.19~10^{-4}\pm7~10^{-6}$&
$-4.34~10^{-4}\pm5~10^{-6}$\\
     $t_E$(days)&$76.8\pm0.1$&$80.6\pm1.5$&$77.2\pm1.3$&$74.9\pm0.9$\\
     $\rho$($\theta_E$)&$5.22~10^{-4}\pm1~10^{-6}$&$4.97~10^{-4}\pm4~10^{-5}$&$5.20~10^{-4}\pm9~10^{-6}$&$5.40~10^{-4}\pm7~10^{-6}$\\
     $I_s$(mag)&$19.00\pm0.01$&$19.05\pm0.02$&$19.01\pm0.02$&$18.97\pm0.01$\\
     $V_s$(mag)&$20.49\pm0.01$&$20.55\pm0.02$&$20.50\pm0.02$&$20.46\pm0.01$\\
     $I_b$(mag)&$18.21\pm0.01$&$18.19\pm0.01$&$18.21\pm0.01$&$18.23\pm0.01$\\
     $V_b$(mag)&$22.72\pm0.10$&$22.35\pm0.10$&$22.70\pm0.11$&$22.98\pm0.11$\\
     $\Pi_{EN}$&&$0.34\pm0.12$&$0.05\pm0.01$&\\
     $\Pi_{EE}$&&$0.28\pm0.04$&$-0.00\pm0.01$&\\
     $s$($\theta_E$)&&&&$1.50547\pm0.04$\\
     $q$&&&&$2.304~10^{-5}\pm1.9~10^{-6}$\\
     $\alpha$(rad)&&&&$-2.39\pm0.02$\\
     $\chi^2$&3647.999&3571.000&3625.150&3258.842\\
       & \\
    \hline
  \end{tabular}
  \centering
  \caption{Model parameters after correction of systematics.}
  \label{tab:modelscorr}
\end{table*}

\begin{table*}
  \centering
  \footnotesize
  \begin{tabular}{lcccccccc}
    & \\
     \hline\hline
     Telescope&RMS&\multicolumn{3}{c}{Raw data}&RMS&\multicolumn{3}{c}{Corrected} \\
	       &(mag) &$\chi^2_{FSPL}$ &$\chi^2_{Pla}$&$\Delta\chi^2 $  &(mag) &$\chi^2_{FSPL} $&$\chi^2_{Pla}$&$\Delta\chi^2$  \\
     \hline
      & \\
      OGLE\_I&0.029&770.308(459)&655.364(456)&114.944&0.029&765.677(459)&661.426(456)&104.251\\
	OGLE\_V&1.348&23.998(20)&24.005(17)&-0.007&1.348&23.998(20)&23.979(17)&0.019\\
      Canopus\_I&0.007&191.996(128)&132.410(125)&59.586&0.007&204.40(128)&137.769(125)&66.631\\
      \textbf{Auckland\_R}&\textbf{0.006}&\textbf{142.839(103)}&\textbf{128.294(100)}&\textbf{14.545}&\textbf{0.006}&\textbf{112.917(103)}&\textbf{72.331(100)}&\textbf{40.586}\\
      \textbf{LSCB\_i}&\textbf{0.007}&\textbf{446.170(374)}&\textbf{317.883(371)}&\textbf{128.287}&\textbf{0.007}&\textbf{442.338(374)}&\textbf{311.049(371)}&\textbf{131.189}\\
      \textbf{LSCA\_i}&\textbf{0.015}&\textbf{399.723(381)}&\textbf{379.065(378)}&\textbf{20.658}&\textbf{0.013}&\textbf{173.683(381)}&\textbf{133.422(378)}&\textbf{40.261}\\
 	CPTA\_i&0.023&21.973(18)&21.926(15)&0.047&0.023&21.972(18)&21.925(15)&0.047\\
	CTIO\_I&0.006&159.174(108)&171.347(15)&-12.173&0.006&158.986(108)&167.376(105)&-8.48\\
	CTIO\_V&0.005&2.986(9)&5.385(6)&-2.399&0.005&3.036(9)&5.227(6)&-2.191\\
	Danish\_z&0.017&448.804(448)&470.960(445)&-22.156&0.017&450.589(448)&471.301(445)&-20.712\\
	MOA\_Red&0.785&727.979(450)&725.116(447)&2.863&0.783&727.760(450)&721.763(447)&5.997\\
	Possum\_N&0.012&355.767(240)&312.101(237)&43.666&0.012&354.002(240)&324.301(237)&29.701\\
	Salerno\_I&0.020&20.016(16)&18.911
(13)&1.105&0.020&19.990(16)&18.872(13)&1.118\\
	Turitea\_R&0.005&29.565(27)&29.341(24)&0.224&0.005&29.539(27)&28.965(24)&0.574\\
	Weizmann\_I&0.034&92.383(56)&92.667(53)&-0.284&0.034&92.389(56)&92.649(53)&-0.260\\
	SAAO\_I&0.014&67.100(54)&66.442(51)&0.658&0.014&67.085(54)&66
.487(51)&0.598\\
      &\\
    \hline
	&\\
	Total &&3900.781&3551.217&349.564&&3647.99&3258.842&389.157\\
	&\\
  \end{tabular}
  \centering
  \caption{$\chi^2$ and RMS of FSPL residuals for each data set before and after correction of systematics. The three corrected data sets are {\bf rendered} in bold. Numbers in parentheses are the degrees of freedom for each model/data set.}
  \label{tab:chi2corr}
\end{table*}
As can be seen in Table~~\ref{tab:chi2corr}, the correction of the systematics  has a significant impact on the LSCA\_i and Auckland\_R data sets, which appear to suffer the most systematics. Note also that the planetary model is more significant after systematics correction, especially for these two telescopes. Even though the planetary model changes slightly before and after correction (the new $s$ and $q$ values are outside the error bars of the uncorrected data sets model), the caustic crossing is virtually unchanged (e.g the central caustic is similar). However, the
clearest signature of the planetary anomaly is still in the Auckland\_R dataset around $HJD\sim6446.02$.

\subsection{Discussion} \label{sec:sysgeneral}
Due to strong finite source effects around a very small central caustic, the suspected planetary signature in OGLE-2013-BLG-0446 is very small. First of all, the low $\chi^2$ improvement ($\Delta\chi^2\sim350$ and $\Delta\chi^2\sim389$ before and after the sysematics correction respectively) of the planetary model is far from the minimum value generally adopted in microlensing for a safe detection \cite{Yee2013}. Note also that even though the caustic crossing is similar, the two planetary models are not fully equivalent. As defined by \cite{Chung2005}, the $R_c$ parameter is the ratio of the vertical length and the horizontal length of a central caustic. This caustic parameter before and after correction is significantly different ($\sim30\%$). Secondly, the highest $\Delta\chi^2$ contributor (LSCB\_i) presents photometric systematics at the same level as the planetary deviations (2 mmag versus 6 mmag). As can be seen in Figure~~\ref{fig:Auckland_corr}, the systematics correction decreases the amplitude of the 'anomaly' in the Auckland\_R dataset and it is therefore better fit by the planetary model. However, the increase in the FSPL residuals after $HJD\sim6446.02$ (from 1\% to zero) is not explained by the planetary model, but similar behaviour is seen in other bright stars. This clearly indicates that bright stars suffer from systematic effects in this dataset which were not revealed by the different quantities we studied.

The planetary model is highly favoured by the Canopus\_I dataset ($\Delta\chi^2\sim60$). However, a closer look at Figure~~\ref{fig:residuals} reveals that the planetary deviations are at a very low level ($\le0.5 \%$). It can not be excluded that the FSPL model correctly fits this data set and that this telescope also suffers from low level systematics errors. We however decided to not realize the same study of photometric systematic errors for the Canopus\_I dataset because there is no obvious deviations in the FSPL residuals and also because enough doubt has already been place in the planetary model we found.

All these points reveal strong doubts about the reality of the planetary signature in OGLE-2013-BLG-0446. Even if we cannot firmly guarantee that the planet is not detected, we prefer to stay conservative and claim that we do not detect a planet in this event. 

\section{Conclusions} \label{sec:conclusions}
We presented the analysis of microlensing event OGLE-2013-BLG-0446. For this highly magnified event ($A\sim3000$), several higher-order effects were investigated in the modeling process: annual and terrestrial parallax and planetary deviations. The study of photometric systematics for several data sets leads to various levels of confidence in the photometry. Moreover, a closer look at the data residuals and a precise study of photometric systematics reveals enough doubt to question any potential signals. Regarding the level of planetary signal ($\sim1\%$) versus the various levels of systematics, we are not confident about the planetary signature in OGLE-2013-BLG-0446. Unfortunately, the clearest signature of the planetary signal was observed only in a single dataset which presents some unexplained behaviour for the brightest stars at the time of the anomaly. These doubts in addition to the relatively low improvement in $\chi^2$ ($\Delta\chi^2\le400$) encourage us to remain conservative and not to claim a planetary detection. This study stresses the importance of studying and quantifying the photometric systematic errors down to the level of 1 \% or lower for the detecion of the smallest microlensing planets.


\begin{acknowledgements}

The authors would like to thank the unknown referee for the usefull comments. This publication was made possible by the NPRP grant \# X-019-1-006
from the Qatar National Research Fund (a member of Qatar Foundation). The statements made herein are
solely the responsibility of the authors. The authors thank the OGLE collaboration for the access of the optimized photometry. This work makes use of observations from the LCOGT network. This research has made use of the LCOGT Archive, which is operated by the California Institute of Technology, under contract with the Las Cumbres Observatory. This research has made much use of NASA's Astrophysics Data System. is research made use of the SIMBAD database
, the VizieR
catalogue access tool, and the cross-match service provide
d by CDS, Strasbourg, France. TS acknowledges the financial support from the JSPS, JSPS23103002,JSPS24253004 and JSPS26247023.
The MOA project is supported by the grant JSPS25103508 and 23340064.
DPB acknowledges support from NSF grants AST-1009621 and AST-1211875,
as well as NASA grants NNX12AF54G and NNX13AF64G. 
Work by IAB and PY was supported by the Marsden Fund of the Royal Society of New Zealand, contract no. MAU1104. 
The operation of the Danish 1.54 m telescope is financed by a grant to UGJ from the Danish Natural Science Research Council. We also acknowledge support from the Center of Excellence Centre for Star and Planet Formation (StarPlan) funded by The Danish National Research Foundation. The MiNDSTEp monitoring campaign is powered by ARTEMiS (Automated Terrestrial Exoplanet Microlensing Search; Dominik et al. 2008, AN 329, 248). KAA, MD, KH, MH, CS, RAS, and YT are thankful to Qatar National Research Fund (QNRF), member of Qatar Foundation, for support by grant NPRP 09-476-1-078. CS has received funding from the European Union Seventh Framework Programme (FP7/2007-2013) under grant agreement No. 268421. Work by C.H. was supported by Creative Research Initiative Program (2009-0081561) of National Research Foundation of Korea. SHG and XBW would like to thank the financial support from National Natural Science Foundation of China through grants Nos. 10873031 and 11473066. TCH acknowledges support from the Korea Research Council of Fundamental Science \& Technology (KRCF) via the KRCF Young Scientist Research Fellowship Programme and for financial support from KASI travel grant number 2013-9-400-00. HK acknowledges the support from the European Commission under the Marie Curie IEF Programme in FP7. MR acknowledges support from FONDECYT postdoctoral fellowship No. 3120097. OW (FNRS research fellow) and JSurdej acknowledge support from the Communaut{\'e} fran\c{c}aise de Belgique - Actions de recherche concert{\'e}es – Acad{\'e}mie Wallonie-Europe.
The HPC (and/or scientific visualization) resources and services used in this work were provided by the IT Research Computing group in Texas A\&M University at Qatar. IT Research Computing is funded by the Qatar Foundation for Education, Science and Community Development (http://www.qf.org.qa).
\end{acknowledgements}


\bibliographystyle{aa}  
\bibliography{biblio_446.bib}

\begin{thebibliography}{60}
\expandafter\ifx\csname natexlab\endcsname\relax\def\natexlab#1{#1}\fi

\bibitem[{{Aigrain} \& {Pont}(2007)}]{Aigrain2007}
{Aigrain}, S. \& {Pont}, F. 2007, \mnras, 378, 741

\bibitem[{{Albrow} {et~al.}(1999){Albrow}, {Beaulieu}, {Caldwell}, {Dominik},
  {Greenhill}, {Hill}, {Kane}, {Martin}, {Menzies}, {Naber}, {Pel}, {Pollard},
  {Sackett}, {Sahu}, {Vermaak}, {Watson}, {Williams}, {Sahu}, \& {PLANET
  Collaboration}}]{Albrow1999}
{Albrow}, M.~D., {Beaulieu}, J.-P., {Caldwell}, J.~A.~R., {et~al.} 1999, \apj,
  522, 1011

\bibitem[{{Albrow} {et~al.}(2009){Albrow}, {Horne}, {Bramich}, {Fouqu{\'e}},
  {Miller}, {Beaulieu}, {Coutures}, {Menzies}, {Williams}, {Batista},
  {Bennett}, {Brillant}, {Cassan}, {Dieters}, {Prester}, {Donatowicz},
  {Greenhill}, {Kains}, {Kane}, {Kubas}, {Marquette}, {Pollard}, {Sahu},
  {Tsapras}, {Wambsganss}, \& {Zub}}]{Albrow2009}
{Albrow}, M.~D., {Horne}, K., {Bramich}, D.~M., {et~al.} 2009, \mnras, 397,
  2099

\bibitem[{{Bachelet} {et~al.}(2012{\natexlab{a}}){Bachelet}, {Fouqu{\'e}},
  {Han}, {Gould}, {Albrow}, {Beaulieu}, {Bertin}, {Bond}, {Christie},
  {Heyrovsk{\'y}}, {Horne}, {J{\o}rgensen}, {Maoz}, {Mathiasen}, {Matsunaga},
  {McCormick}, {Menzies}, {Nataf}, {Natusch}, {Oi}, {Renon}, {Tsapras},
  {Udalski}, {Yee}, {Batista}, {Bennett}, {Brillant}, {Caldwell}, {Cassan},
  {Cole}, {Cook}, {Coutures}, {Dieters}, {Dominik}, {Dominis Prester},
  {Donatowicz}, {Greenhill}, {Kains}, {Kane}, {Marquette}, {Martin}, {Pollard},
  {Sahu}, {Street}, {Wambsganss}, {Williams}, {Zub}, {PLANET Collaboration},
  {Bos}, {Dong}, {Drummond}, {Gaudi}, {Graff}, {Janczak}, {Kaspi},
  {Koz{\l}owski}, {Lee}, {Monard}, {Mu{\~n}oz}, {Park}, {Pogge}, {Polishook},
  {Shporer}, {Fun Collaboration}, {Abe}, {Botzler}, {Fukui}, {Furusawa},
  {Hearnshaw}, {Itow}, {Korpela}, {Ling}, {Masuda}, {Matsubara}, {Miyake},
  {Muraki}, {Ohnishi}, {Rattenbury}, {Saito}, {Sullivan}, {Sumi}, {Suzuki},
  {Sweatman}, {Tristram}, {Wada}, {Moa Collaboration}, {Allan}, {Bode},
  {Bramich}, {Clay}, {Fraser}, {Hawkins}, {Kerins}, {Lister}, {Mottram},
  {Saunders}, {Snodgrass}, {Steele}, {Wheatley}, {Robonet-Ii Collaboration},
  {Bozza}, {Browne}, {Burgdorf}, {Calchi Novati}, {Dreizler}, {Finet},
  {Glitrup}, {Grundahl}, {Harps{\o}E}, {Hessman}, {Hinse}, {Hundertmark},
  {Liebig}, {Maier}, {Mancini}, {Rahvar}, {Ricci}, {Scarpetta}, {Skottfelt},
  {Southworth}, {Surdej}, {Zimmer}, \& {Mindstep Consortium}}]{Bachelet2012b}
{Bachelet}, E., {Fouqu{\'e}}, P., {Han}, C., {et~al.} 2012{\natexlab{a}}, \aap,
  547, A55

\bibitem[{{Bachelet} {et~al.}(2012{\natexlab{b}}){Bachelet}, {Shin}, {Han},
  {Fouqu{\'e}}, {Gould}, {Menzies}, {Beaulieu}, {Bennett}, {Bond}, {Dong},
  {Heyrovsk{\'y}}, {Marquette}, {Marshall}, {Skowron}, {Street}, {Sumi},
  {Udalski}, {Abe}, {Agabi}, {Albrow}, {Allen}, {Bertin}, {Bos}, {Bramich},
  {Chavez}, {Christie}, {Cole}, {Crouzet}, {Dieters}, {Dominik}, {Drummond},
  {Greenhill}, {Guillot}, {Henderson}, {Hessman}, {Horne}, {Hundertmark},
  {Johnson}, {J{\o}rgensen}, {Kandori}, {Liebig}, {M{\'e}karnia}, {McCormick},
  {Moorhouse}, {Nagayama}, {Nataf}, {Natusch}, {Nishiyama}, {Rivet}, {Sahu},
  {Shvartzvald}, {Thornley}, {Tomczak}, {Tsapras}, {Yee}, {Batista}, {Bennett},
  {Brillant}, {Caldwell}, {Cassan}, {Corrales}, {Coutures}, {Dominis Prester},
  {Donatowicz}, {Kubas}, {Martin}, {Williams}, {Zub}, {The PLANET
  Collaboration}, {de Almeida}, {DePoy}, {Gaudi}, {Hung}, {Jablonski}, {Kaspi},
  {Klein}, {Lee}, {Lee}, {Koo}, {Maoz}, {Mu{\~n}oz}, {Pogge}, {Polishook},
  {Shporer}, {{$\mu$}Collaboration}, {Abe}, {Botzler}, {Chote}, {Freeman},
  {Fukui}, {Furusawa}, {Harris}, {Itow}, {Kobara}, {Ling}, {Masuda},
  {Matsubara}, {Miyake}, {Ohmori}, {Ohnishi}, {Rattenbury}, {Saito},
  {Sullivan}, {Suzuki}, {Sweatman}, {Tristram}, {Wada}, {Yock}, {The MOA
  Collaboration}, {Szyma{\'n}ski}, {Soszy{\'n}ski}, {Kubiak}, {Poleski},
  {Ulaczyk}, {Pietrzy{\'n}ski}, {Wyrzykowski}, {The OGLE Collaboration},
  {Kains}, {Snodgrass}, {Steele}, {The RoboNet Collaboration}, {Alsubai},
  {Bozza}, {Browne}, {Burgdorf}, {Calchi Novati}, {Dodds}, {Dreizler}, {Finet},
  {Gerner}, {Hardis}, {Harps{\o}e}, {Hinse}, {Kerins}, {Mancini}, {Mathiasen},
  {Penny}, {Proft}, {Rahvar}, {Ricci}, {Scarpetta}, {Sch{\"a}fer},
  {Sch{\"o}nebeck}, {Southworth}, {Surdej}, {Wambsganss}, \& {MiNDSTEp
  Consortium}}]{Bachelet2012a}
{Bachelet}, E., {Shin}, I.-G., {Han}, C., {et~al.} 2012{\natexlab{b}}, \apj,
  754, 73

\bibitem[{{Beaulieu} {et~al.}(2006){Beaulieu}, {Bennett}, {Fouqu{\'e}},
  {Williams}, {Dominik}, {J{\o}rgensen}, {Kubas}, {Cassan}, {Coutures},
  {Greenhill}, {Hill}, {Menzies}, {Sackett}, {Albrow}, {Brillant}, {Caldwell},
  {Calitz}, {Cook}, {Corrales}, {Desort}, {Dieters}, {Dominis}, {Donatowicz},
  {Hoffman}, {Kane}, {Marquette}, {Martin}, {Meintjes}, {Pollard}, {Sahu},
  {Vinter}, {Wambsganss}, {Woller}, {Horne}, {Steele}, {Bramich}, {Burgdorf},
  {Snodgrass}, {Bode}, {Udalski}, {Szyma{\'n}ski}, {Kubiak}, {Wi{\c e}ckowski},
  {Pietrzy{\'n}ski}, {Soszy{\'n}ski}, {Szewczyk}, {Wyrzykowski},
  {Paczy{\'n}ski}, {Abe}, {Bond}, {Britton}, {Gilmore}, {Hearnshaw}, {Itow},
  {Kamiya}, {Kilmartin}, {Korpela}, {Masuda}, {Matsubara}, {Motomura},
  {Muraki}, {Nakamura}, {Okada}, {Ohnishi}, {Rattenbury}, {Sako}, {Sato},
  {Sasaki}, {Sekiguchi}, {Sullivan}, {Tristram}, {Yock}, \&
  {Yoshioka}}]{Beaulieu2006}
{Beaulieu}, J.-P., {Bennett}, D.~P., {Fouqu{\'e}}, P., {et~al.} 2006, \nat,
  439, 437

\bibitem[{{Bennett} {et~al.}(2014){Bennett}, {Batista}, {Bond}, {Bennett},
  {Suzuki}, {Beaulieu}, {Udalski}, {Donatowicz}, {Bozza}, {Abe}, {Botzler},
  {Freeman}, {Fukunaga}, {Fukui}, {Itow}, {Koshimoto}, {Ling}, {Masuda},
  {Matsubara}, {Muraki}, {Namba}, {Ohnishi}, {Rattenbury}, {Saito}, {Sullivan},
  {Sumi}, {Sweatman}, {Tristram}, {Tsurumi}, {Wada}, {Yock}, {MOA
  Collaboration}, {Albrow}, {Bachelet}, {Brillant}, {Caldwell}, {Cassan},
  {Cole}, {Corrales}, {Coutures}, {Dieters}, {Dominis Prester}, {Fouqu{\'e}},
  {Greenhill}, {Horne}, {Koo}, {Kubas}, {Marquette}, {Martin}, {Menzies},
  {Sahu}, {Wambsganss}, {Williams}, {Zub}, {PLANET Collaboration}, {Choi},
  {DePoy}, {Dong}, {Gaudi}, {Gould}, {Han}, {Henderson}, {McGregor}, {Lee},
  {Pogge}, {Shin}, {Yee}, {The {$\mu$}FUN Collaboration}, {Szyma{\'n}ski},
  {Skowron}, {Poleski}, {Koz{\l}owski}, {Wyrzykowski}, {Kubiak},
  {Pietrukowicz}, {Pietrzy{\'n}ski}, {Soszy{\'n}ski}, {Ulaczyk}, {The OGLE
  Collaboration}, {Tsapras}, {Street}, {Dominik}, {Bramich}, {Browne},
  {Hundertmark}, {Kains}, {Snodgrass}, {Steele}, {The RoboNet Collaboration},
  {Dekany}, {Gonzalez}, {Heyrovsk{\'y}}, {Kandori}, {Kerins}, {Lucas},
  {Minniti}, {Nagayama}, {Rejkuba}, {Robin}, \& {Saito}}]{Bennett2014}
{Bennett}, D.~P., {Batista}, V., {Bond}, I.~A., {et~al.} 2014, \apj, 785, 155

\bibitem[{{Bennett} \& {Rhie}(1996)}]{Bennett1996}
{Bennett}, D.~P. \& {Rhie}, S.~H. 1996, \apj, 472, 660

\bibitem[{{Bessell} \& {Brett}(1988)}]{Bessell1988}
{Bessell}, M.~S. \& {Brett}, J.~M. 1988, \pasp, 100, 1134

\bibitem[{{Bond} {et~al.}(2001){Bond}, {Abe}, {Dodd}, {Hearnshaw}, {Honda},
  {Jugaku}, {Kilmartin}, {Marles}, {Masuda}, {Matsubara}, {Muraki}, {Nakamura},
  {Nankivell}, {Noda}, {Noguchi}, {Ohnishi}, {Rattenbury}, {Reid}, {Saito},
  {Sato}, {Sekiguchi}, {Skuljan}, {Sullivan}, {Sumi}, {Takeuti}, {Watase},
  {Wilkinson}, {Yamada}, {Yanagisawa}, \& {Yock}}]{Bond2001}
{Bond}, I.~A., {Abe}, F., {Dodd}, R.~J., {et~al.} 2001, \mnras, 327, 868

\bibitem[{{Bramich}(2008)}]{Bramich2008}
{Bramich}, D.~M. 2008, \mnras, 386, L77

\bibitem[{{Bramich} {et~al.}(2015){Bramich}, {Bachelet}, {Alsubai}, {Mislis},
  \& {Parley}}]{Bramich2015}
{Bramich}, D.~M., {Bachelet}, E., {Alsubai}, K.~A., {Mislis}, D., \& {Parley},
  N. 2015, \aap, accepted

\bibitem[{{Bramich} \& {Freudling}(2012)}]{Bramich2012}
{Bramich}, D.~M. \& {Freudling}, W. 2012, \mnras, 424, 1584

\bibitem[{{Bramich} {et~al.}(2013){Bramich}, {Horne}, {Albrow}, {Tsapras},
  {Snodgrass}, {Street}, {Hundertmark}, {Kains}, {Arellano Ferro}, {Figuera},
  \& {Giridhar}}]{Bramich2013}
{Bramich}, D.~M., {Horne}, K., {Albrow}, M.~D., {et~al.} 2013, \mnras, 428,
  2275

\bibitem[{{Casagrande} {et~al.}(2010){Casagrande}, {Ram{\'{\i}}rez},
  {Mel{\'e}ndez}, {Bessell}, \& {Asplund}}]{Casagrande2010}
{Casagrande}, L., {Ram{\'{\i}}rez}, I., {Mel{\'e}ndez}, J., {Bessell}, M., \&
  {Asplund}, M. 2010, \aap, 512, A54

\bibitem[{{Chung} {et~al.}(2005){Chung}, {Han}, {Park}, {Kim}, {Kang}, {Ryu},
  {Kim}, {Jeon}, {Lee}, {Chang}, {Lee}, \& {Kang}}]{Chung2005}
{Chung}, S.-J., {Han}, C., {Park}, B.-G., {et~al.} 2005, \apj, 630, 535

\bibitem[{{Claret}(2000)}]{Claret2000}
{Claret}, A. 2000, \aap, 363, 1081

\bibitem[{{Coccato} {et~al.}(2014){Coccato}, {Bramich}, {Freudling}, \&
  {Moehler}}]{Coccato2014}
{Coccato}, L., {Bramich}, D.~M., {Freudling}, W., \& {Moehler}, S. 2014,
  \mnras, 438, 1256

\bibitem[{{Dominik} {et~al.}(2008){Dominik}, {Horne}, {Allan}, {Rattenbury},
  {Tsapras}, {Snodgrass}, {Bode}, {Burgdorf}, {Fraser}, {Kerins}, {Mottram},
  {Steele}, {Street}, {Wheatley}, \& {Wyrzykowski}}]{Dominik2008}
{Dominik}, M., {Horne}, K., {Allan}, A., {et~al.} 2008, Astronomische
  Nachrichten, 329, 248

\bibitem[{{Dong} {et~al.}(2009){Dong}, {Bond}, {Gould}, {Koz{\l}owski},
  {Miyake}, {Gaudi}, {Bennett}, {Abe}, {Gilmore}, {Fukui}, {Furusawa},
  {Hearnshaw}, {Itow}, {Kamiya}, {Kilmartin}, {Korpela}, {Lin}, {Ling},
  {Masuda}, {Matsubara}, {Muraki}, {Nagaya}, {Ohnishi}, {Okumura}, {Perrott},
  {Rattenbury}, {Saito}, {Sako}, {Sato}, {Skuljan}, {Sullivan}, {Sumi},
  {Sweatman}, {Tristram}, {Yock}, {The MOA Collaboration}, {Bolt}, {Christie},
  {DePoy}, {Han}, {Janczak}, {Lee}, {Mallia}, {McCormick}, {Monard}, {Maury},
  {Natusch}, {Park}, {Pogge}, {Santallo}, {Stanek}, {The {$\mu$}FUN
  Collaboration}, {Udalski}, {Kubiak}, {Szyma{\'n}ski}, {Pietrzy{\'n}ski},
  {Soszy{\'n}ski}, {Szewczyk}, {Wyrzykowski}, {Ulaczyk}, \& {The OGLE
  Collaboration}}]{Dong2009b}
{Dong}, S., {Bond}, I.~A., {Gould}, A., {et~al.} 2009, \apj, 698, 1826

\bibitem[{{Dong} {et~al.}(2007){Dong}, {Udalski}, {Gould}, {Reach}, {Christie},
  {Boden}, {Bennett}, {Fazio}, {Griest}, {Szyma{\'n}ski}, {Kubiak},
  {Soszy{\'n}ski}, {Pietrzy{\'n}ski}, {Szewczyk}, {Wyrzykowski}, {Ulaczyk},
  {Wieckowski}, {Paczy{\'n}ski}, {DePoy}, {Pogge}, {Preston}, {Thompson}, \&
  {Patten}}]{Dong2007}
{Dong}, S., {Udalski}, A., {Gould}, A., {et~al.} 2007, \apj, 664, 862

\bibitem[{{Gould}(1994)}]{Gould1994}
{Gould}, A. 1994, \apjl, 421, L71

\bibitem[{{Gould}(1997)}]{Gould1997}
{Gould}, A. 1997, \apj, 480, 188

\bibitem[{{Gould}(2000)}]{Gould2000}
{Gould}, A. 2000, \apj, 542, 785

\bibitem[{{Gould}(2004)}]{Gould2004}
{Gould}, A. 2004, \apj, 606, 319

\bibitem[{{Gould} {et~al.}(2009){Gould}, {Dong}, {Bennett}, {Bond}, {Udalski},
  \& {Kozlowski}}]{Gould2010}
{Gould}, A., {Dong}, S., {Bennett}, D.~P., {et~al.} 2009, arXiv0910.1832

\bibitem[{{Gould} \& {Loeb}(1992)}]{Gould1992}
{Gould}, A. \& {Loeb}, A. 1992, \apj, 396, 104

\bibitem[{{Gould} {et~al.}(2006){Gould}, {Udalski}, {An}, {Bennett}, {Zhou},
  {Dong}, {Rattenbury}, {Gaudi}, {Yock}, {Bond}, {Christie}, {Horne},
  {Anderson}, {Stanek}, {DePoy}, {Han}, {McCormick}, {Park}, {Pogge},
  {Poindexter}, {Soszy{\'n}ski}, {Szyma{\'n}ski}, {Kubiak}, {Pietrzy{\'n}ski},
  {Szewczyk}, {Wyrzykowski}, {Ulaczyk}, {Paczy{\'n}ski}, {Bramich},
  {Snodgrass}, {Steele}, {Burgdorf}, {Bode}, {Botzler}, {Mao}, \&
  {Swaving}}]{Gould2006}
{Gould}, A., {Udalski}, A., {An}, D., {et~al.} 2006, \apjl, 644, L37

\bibitem[{{Gould} \& {Yee}(2013)}]{Gould2013}
{Gould}, A. \& {Yee}, J.~C. 2013, \apj, 764, 107

\bibitem[{{Griest} \& {Safizadeh}(1998)}]{Griest1998}
{Griest}, K. \& {Safizadeh}, N. 1998, \apj, 500, 37

\bibitem[{{Hardy} \& {Walker}(1995)}]{Hardy1995}
{Hardy}, S.~J. \& {Walker}, M.~A. 1995, \mnras, 276, L79

\bibitem[{{Holz} \& {Wald}(1996)}]{Holz1996}
{Holz}, D.~E. \& {Wald}, R.~M. 1996, \apj, 471, 64

\bibitem[{{Honeycutt}(1992)}]{Honeycutt1992}
{Honeycutt}, R.~K. 1992, \pasp, 104, 435

\bibitem[{Kains(2015)}]{Kains2015}
Kains, N. 2015, submitted

\bibitem[{{Kervella} \& {Fouqu{\'e}}(2008)}]{Kervella2008}
{Kervella}, P. \& {Fouqu{\'e}}, P. 2008, \aap, 491, 855

\bibitem[{{Kov{\'a}cs} {et~al.}(2005){Kov{\'a}cs}, {Bakos}, \&
  {Noyes}}]{Kovacs2005}
{Kov{\'a}cs}, G., {Bakos}, G., \& {Noyes}, R.~W. 2005, \mnras, 356, 557

\bibitem[{{Mall{\'e}n-Ornelas} {et~al.}(2003){Mall{\'e}n-Ornelas}, {Seager},
  {Yee}, {Minniti}, {Gladders}, {Mall{\'e}n-Fullerton}, \&
  {Brown}}]{Mallen2003}
{Mall{\'e}n-Ornelas}, G., {Seager}, S., {Yee}, H.~K.~C., {et~al.} 2003, \apj,
  582, 1123

\bibitem[{{Milne}(1921)}]{Milne1921}
{Milne}, E.~A. 1921, \mnras, 81, 361

\bibitem[{{Miyake} {et~al.}(2012){Miyake}, {Udalski}, {Sumi}, {Bennett},
  {Dong}, {Street}, {Greenhill}, {Bond}, {Gould}, {Kubiak}, {Szyma{\'n}ski},
  {Pietrzy{\'n}ski}, {Soszy{\'n}ski}, {Ulaczyk}, {Wyrzykowski}, {OGLE
  Collaboration}, {Abe}, {Fukui}, {Furusawa}, {Holderness}, {Itow}, {Korpela},
  {Ling}, {Masuda}, {Matsubara}, {Muraki}, {Nagayama}, {Ohnishi}, {Rattenbury},
  {Saito}, {Sako}, {Sullivan}, {Sweatman}, {Tristram}, {Yock}, {MOA
  Collaboration}, {Allen}, {Christie}, {DePoy}, {Gaudi}, {Han}, {Lee},
  {McCormick}, {Monard}, {Natusch}, {Park}, {Pogge}, {{$\mu$}FUN
  Collaboration}, {Allan}, {Bode}, {Bramich}, {Clay}, {Dominik}, {Horne},
  {Kains}, {Mottram}, {Snodgrass}, {Steele}, {Tsapras}, {RoboNet
  Collaboration}, {Albrow}, {Batista}, {Beaulieu}, {Brillant}, {Burgdorf},
  {Caldwell}, {Cassan}, {Cole}, {Cook}, {Coutures}, {Dieters}, {Dominis
  Prester}, {Donatowicz}, {Fouqu{\'e}}, {Jorgensen}, {Kane}, {Kubas},
  {Marquette}, {Martin}, {Menzies}, {Pollard}, {Sahu}, {Wambsganss},
  {Williams}, {Zub}, \& {PLANET Collaboration}}]{Miyake2012}
{Miyake}, N., {Udalski}, A., {Sumi}, T., {et~al.} 2012, \apj, 752, 82

\bibitem[{{Nataf} {et~al.}(2013){Nataf}, {Gould}, {Fouqu{\'e}}, {Gonzalez},
  {Johnson}, {Skowron}, {Udalski}, {Szyma{\'n}ski}, {Kubiak},
  {Pietrzy{\'n}ski}, {Soszy{\'n}ski}, {Ulaczyk}, {Wyrzykowski}, \&
  {Poleski}}]{Nataf2012}
{Nataf}, D.~M., {Gould}, A., {Fouqu{\'e}}, P., {et~al.} 2013, \apj, 769, 88

\bibitem[{{Nemiroff} \& {Wickramasinghe}(1994)}]{Nemiroff1994}
{Nemiroff}, R.~J. \& {Wickramasinghe}, W.~A.~D.~T. 1994, \apjl, 424, L21

\bibitem[{{Paczy\'nski}(1986)}]{Paczynski1986}
{Paczy\'nski}, B. 1986, \apj, 304, 1

\bibitem[{{Padmanabhan} {et~al.}(2008){Padmanabhan}, {Schlegel}, {Finkbeiner},
  {Barentine}, {Blanton}, {Brewington}, {Gunn}, {Harvanek}, {Hogg},
  {Ivezi{\'c}}, {Johnston}, {Kent}, {Kleinman}, {Knapp}, {Krzesinski}, {Long},
  {Neilsen}, {Nitta}, {Loomis}, {Lupton}, {Roweis}, {Snedden}, {Strauss}, \&
  {Tucker}}]{Padmanabhan2008}
{Padmanabhan}, N., {Schlegel}, D.~J., {Finkbeiner}, D.~P., {et~al.} 2008, \apj,
  674, 1217

\bibitem[{{Pepper} \& {Gaudi}(2005)}]{Pepper2005}
{Pepper}, J. \& {Gaudi}, B.~S. 2005, \apj, 631, 581

\bibitem[{{Pont} {et~al.}(2006){Pont}, {Zucker}, \& {Queloz}}]{Pont2006}
{Pont}, F., {Zucker}, S., \& {Queloz}, D. 2006, \mnras, 373, 231

\bibitem[{{Regnault} {et~al.}(2009){Regnault}, {Conley}, {Guy}, {Sullivan},
  {Cuillandre}, {Astier}, {Balland}, {Basa}, {Carlberg}, {Fouchez}, {Hardin},
  {Hook}, {Howell}, {Pain}, {Perrett}, \& {Pritchet}}]{Regnault2009}
{Regnault}, N., {Conley}, A., {Guy}, J., {et~al.} 2009, \aap, 506, 999

\bibitem[{{Skottfelt} {et~al.}(2015){Skottfelt}, {Bramich}, {Hundertmark},
  {J{\o}rgensen}, {Michaelsen}, {Kj{\ae}rgaard}, {Southworth}, {S{\o}rensen},
  {Andersen}, {Andersen}, {Christensen-Dalsgaard}, {Frandsen}, {Grundahl},
  {Harps{\o}e}, {Kjeldsen}, \& {Pall{\'e}}}]{Skottfelt2015}
{Skottfelt}, J., {Bramich}, D.~M., {Hundertmark}, M., {et~al.} 2015, \aap, 574,
  A54

\bibitem[{{Skowron} {et~al.}(2011){Skowron}, {Udalski}, {Gould}, {Dong},
  {Monard}, {Han}, {Nelson}, {McCormick}, {Moorhouse}, {Thornley}, {Maury},
  {Bramich}, {Greenhill}, {Koz{\l}owski}, {Bond}, {Poleski}, {Wyrzykowski},
  {Ulaczyk}, {Kubiak}, {Szyma{\'n}ski}, {Pietrzy{\'n}ski}, {Soszy{\'n}ski},
  {OGLE Collaboration}, {Gaudi}, {Yee}, {Hung}, {Pogge}, {DePoy}, {Lee},
  {Park}, {Allen}, {Mallia}, {Drummond}, {Bolt}, {{$\mu$}FUN Collaboration},
  {Allan}, {Browne}, {Clay}, {Dominik}, {Fraser}, {Horne}, {Kains}, {Mottram},
  {Snodgrass}, {Steele}, {Street}, {Tsapras}, {RoboNet Collaboration}, {Abe},
  {Bennett}, {Botzler}, {Douchin}, {Freeman}, {Fukui}, {Furusawa}, {Hayashi},
  {Hearnshaw}, {Hosaka}, {Itow}, {Kamiya}, {Kilmartin}, {Korpela}, {Lin},
  {Ling}, {Makita}, {Masuda}, {Matsubara}, {Muraki}, {Nagayama}, {Miyake},
  {Nishimoto}, {Ohnishi}, {Perrott}, {Rattenbury}, {Saito}, {Skuljan},
  {Sullivan}, {Sumi}, {Suzuki}, {Sweatman}, {Tristram}, {Wada}, {Yock}, {MOA
  Collaboration}, {Beaulieu}, {Fouqu{\'e}}, {Albrow}, {Batista}, {Brillant},
  {Caldwell}, {Cassan}, {Cole}, {Cook}, {Coutures}, {Dieters}, {Dominis
  Prester}, {Donatowicz}, {Kane}, {Kubas}, {Marquette}, {Martin}, {Menzies},
  {Sahu}, {Wambsganss}, {Williams}, {Zub}, \& {PLANET
  Collaboration}}]{Skowron2011}
{Skowron}, J., {Udalski}, A., {Gould}, A., {et~al.} 2011, \apj, 738, 87

\bibitem[{{Smith} {et~al.}(2012){Smith}, {Stumpe}, {Van Cleve}, {Jenkins},
  {Barclay}, {Fanelli}, {Girouard}, {Kolodziejczak}, {McCauliff}, {Morris}, \&
  {Twicken}}]{Smith2012}
{Smith}, J.~C., {Stumpe}, M.~C., {Van Cleve}, J.~E., {et~al.} 2012, \pasp, 124,
  1000

\bibitem[{{Smith} {et~al.}(2003){Smith}, {Mao}, \& {Paczy{\'n}ski}}]{Smith2003}
{Smith}, M.~C., {Mao}, S., \& {Paczy{\'n}ski}, B. 2003, \mnras, 339, 925

\bibitem[{{Tamuz} {et~al.}(2005){Tamuz}, {Mazeh}, \& {Zucker}}]{Tamuz2005}
{Tamuz}, O., {Mazeh}, T., \& {Zucker}, S. 2005, \mnras, 356, 1466

\bibitem[{{Tsapras} {et~al.}(2009){Tsapras}, {Street}, {Horne}, {Snodgrass},
  {Dominik}, {Allan}, {Steele}, {Bramich}, {Saunders}, {Rattenbury}, {Mottram},
  {Fraser}, {Clay}, {Burgdorf}, {Bode}, {Lister}, {Hawkins}, {Beaulieu},
  {Fouqu{\'e}}, {Albrow}, {Menzies}, {Cassan}, \&
  {Dominis-Prester}}]{Tsapras2009}
{Tsapras}, Y., {Street}, R., {Horne}, K., {et~al.} 2009, Astronomische
  Nachrichten, 330, 4

\bibitem[{{Udalski}(2003{\natexlab{a}})}]{Udalski2003a}
{Udalski}, A. 2003{\natexlab{a}}, \apj, 590, 284

\bibitem[{{Udalski}(2003{\natexlab{b}})}]{Udalski2003b}
{Udalski}, A. 2003{\natexlab{b}}, \actaa, 53, 291

\bibitem[{{Udalski} \& {Szyma{\'n}ski}(2015)}]{Udalski2015}
{Udalski}, A. \& {Szyma{\'n}ski}, M. 2015, \actaa, 65, 1

\bibitem[{{Vermaak}(2000)}]{Vermaak2000}
{Vermaak}, P. 2000, \mnras, 319, 1011

\bibitem[{{Witt} \& {Mao}(1994)}]{Witt1994}
{Witt}, H.~J. \& {Mao}, S. 1994, \apj, 430, 505

\bibitem[{{Yee} {et~al.}(2013){Yee}, {Hung}, {Bond}, {Allen}, {Monard},
  {Albrow}, {Fouqu{\'e}}, {Dominik}, {Tsapras}, {Udalski}, {Gould}, {Zellem},
  {Bos}, {Christie}, {DePoy}, {Dong}, {Drummond}, {Gaudi}, {Gorbikov}, {Han},
  {Kaspi}, {Klein}, {Lee}, {Maoz}, {McCormick}, {Moorhouse}, {Natusch}, {Nola},
  {Park}, {Pogge}, {Polishook}, {Shporer}, {Shvartzvald}, {Skowron},
  {Thornley}, {{$\mu$}FUN Collaboration}, {Abe}, {Bennett}, {Botzler}, {Chote},
  {Freeman}, {Fukui}, {Furusawa}, {Harris}, {Itow}, {Ling}, {Masuda},
  {Matsubara}, {Miyake}, {Ohnishi}, {Rattenbury}, {Saito}, {Sullivan}, {Sumi},
  {Suzuki}, {Sweatman}, {Tristram}, {Wada}, {Yock}, {MOA Collaboration},
  {Szyma{\'n}ski}, {Soszy{\'n}ski}, {Kubiak}, {Poleski}, {Ulaczyk},
  {Pietrzy{\'n}ski}, {Wyrzykowski}, {The OGLE Collaboration}, {Bachelet},
  {Batista}, {Beatty}, {Beaulieu}, {Bennett}, {Bowens-Rubin}, {Brillant},
  {Caldwell}, {Cassan}, {Cole}, {Corrales}, {Coutures}, {Dieters}, {Dominis
  Prester}, {Donatowicz}, {Greenhill}, {Henderson}, {Kubas}, {Marquette},
  {Martin}, {Menzies}, {Shappee}, {Williams}, {Wouters}, {van Saders}, {Zub},
  {The PLANET Collaboration}, {Street}, {Horne}, {Bramich}, {Steele}, {The
  RoboNet Collaboration}, {Alsubai}, {Bozza}, {Browne}, {Burgdorf}, {Calchi
  Novati}, {Dodds}, {Finet}, {Gerner}, {Hardis}, {Harps{\o}e}, {Hessman},
  {Hinse}, {Hundertmark}, {J{\o}rgensen}, {Kains}, {Kerins}, {Liebig},
  {Mancini}, {Mathiasen}, {Penny}, {Proft}, {Rahvar}, {Ricci}, {Sahu},
  {Scarpetta}, {Sch{\"a}fer}, {Sch{\"o}nebeck}, {Snodgrass}, {Southworth},
  {Surdej}, {Wambsganss}, \& {MiNDSTEp Consortium}}]{Yee2013}
{Yee}, J.~C., {Hung}, L.-W., {Bond}, I.~A., {et~al.} 2013, \apj, 769, 77

\bibitem[{{Yee} {et~al.}(2009){Yee}, {Udalski}, {Sumi}, {Dong}, {Koz{\l}owski},
  {Bird}, {Cole}, {Higgins}, {McCormick}, {Monard}, {Polishook}, {Shporer},
  {Spector}, {Szyma{\'n}ski}, {Kubiak}, {Pietrzy{\'n}ski}, {Soszy{\'n}ski},
  {Szewczyk}, {Ulaczyk}, {Wyrzykowski}, {Poleski}, {The OGLE Collaboration},
  {Allen}, {Bos}, {Christie}, {DePoy}, {Eastman}, {Gaudi}, {Gould}, {Han},
  {Kaspi}, {Lee}, {Mallia}, {Maury}, {Maoz}, {Natusch}, {Park}, {Pogge},
  {Santallo}, {The {$\mu$}FUN Collaboration}, {Abe}, {Bond}, {Fukui},
  {Furusawa}, {Hearnshaw}, {Hosaka}, {Itow}, {Kamiya}, {Korpela}, {Kilmartin},
  {Lin}, {Ling}, {Makita}, {Masuda}, {Matsubara}, {Miyake}, {Muraki}, {Nagaya},
  {Nishimoto}, {Ohnishi}, {Perrott}, {Rattenbury}, {Sako}, {Saito}, {Skuljan},
  {Sullivan}, {Sweatman}, {Tristram}, {Yock}, {The MOA Collaboration},
  {Albrow}, {Batista}, {Fouqu{\'e}}, {Beaulieu}, {Bennett}, {Cassan},
  {Comparat}, {Coutures}, {Dieters}, {Greenhill}, {Horne}, {Kains}, {Kubas},
  {Martin}, {Menzies}, {Wambsganss}, {Williams}, {Zub}, \& {The PLANET
  Collaboration}}]{Yee2009}
{Yee}, J.~C., {Udalski}, A., {Sumi}, T., {et~al.} 2009, \apj, 703, 2082

\bibitem[{{Yoo} {et~al.}(2004){Yoo}, {DePoy}, {Gal-Yam}, {Gaudi}, {Gould},
  {Han}, {Lipkin}, {Maoz}, {Ofek}, {Park}, {Pogge}, {Udalski}, {Soszy{\'
  n}ski}, {Wyrzykowski}, {Kubiak}, {Szyma{\' n}ski}, {Pietrzy{\' n}ski},
  {Szewczyk}, \& {{\. Z}ebru{\' n}}}]{Yoo2004}
{Yoo}, J., {DePoy}, D.~L., {Gal-Yam}, A., {et~al.} 2004, \apj, 603, 139

\end{thebibliography}
\end{document}